\documentclass[12pt,a4paper]{amsart}

\usepackage{empheq}
\usepackage{tcolorbox}
\usepackage{framed}
\usepackage{amsmath}
\usepackage[latin9]{inputenc}
\usepackage{amstext}
\usepackage{amssymb}
\usepackage{caption}
\usepackage{longtable}
\usepackage{lscape}
\usepackage{tabularx}
\usepackage{varioref}
\usepackage{amsthm}
\usepackage{graphicx}
\usepackage{txfonts}
\usepackage{pxfonts}
\usepackage{marginnote}
\usepackage{epic}
\usepackage{eepic}
\usepackage{float}
\usepackage{rotating}
\usepackage{epsfig}
\usepackage{indentfirst}
\usepackage{array}
\usepackage{varioref}
\usepackage{tikz}
\usepackage{marginnote}
\usepackage{pgf}
\usepackage{bbm}
\usepackage{appendix}
\usepackage{amsbsy}
\usepackage{latexsym}
\usepackage{amsfonts}
\usepackage{pstricks}
\usepackage{color}
\usepackage[numbers,square,comma, compress]{natbib}
\usepackage{eurosym}
\usepackage{tikz}
\usepackage{pdfpages}
\usepackage{wasysym}
\usepackage{caption}
\usepackage{subcaption}
\usepackage{braket}
\usepackage{simplewick}

\usepackage[scale=0.82]{geometry}

\makeatletter

\pdfpageheight\paperheight
\pdfpagewidth\paperwidth

\numberwithin{equation}{section}
\numberwithin{figure}{section}

\numberwithin{equation}{section}
\numberwithin{figure}{section}

\let\emptyset\varnothing

\def\bW{\bar{W}}

\def\ll{\left\lgroup}
\def\rr{\right\rgroup}

\def\leq{\leqslant}
\def\geq{\geqslant}

\def\bab{\pmb{b}}

\def\bx{\bar{x}}
\def\by{\bar{y}}

\def\i{\iota}

\def\x{x^{(1)}}

\def\bref\textbf{\ref}

\def\ll{ \left\lgroup}
\def\rr{\right\rgroup}

\newcommand{\cC}{\mathcal{C}}

\newcommand{\cE}{\mathcal{E}}

\newcommand{\cM}{\mathcal{M}}

\newcommand{\cO}{\mathcal{O}}

\newcommand{\cR}{\mathcal{R}}

\newcommand{\ZZ}{\mathbbm{Z}}

\def\bi{\pmb{\iota }}
\def\bj{\pmb{\jmath}}

\newcommand{\1}{\textbf{ 1.}}
\newcommand{\2}{\textbf{ 2.}}
\newcommand{\3}{\textbf{ 3.}}
\newcommand{\4}{\textbf{ 4.}}
\newcommand{\5}{\textbf{ 5.}}

\newcommand{\wsq}{\square}

\def\ba{\pmb{a}}
\def\bfq{\pmb{p}}
\def\bfq{\pmb{q}}
\def\bx{\pmb{x}}
\def\by{\pmb{y}}

\def\bR{\pmb{R}}
\def\bS{\pmb{S}}

\def\bW{\pmb{W}}
\def\bX{\pmb{X}}
\def\bY{\pmb{Y}}

\def\bi{\pmb{\iota }}
\def\bj{\pmb{\jmath}} 

\def\i{\iota}

\restylefloat{figure}

\usepackage[parfill]{parskip}

\begin{document} 

\begin{flushright}
    UT-18-07
\end{flushright}
	
\title[]{
An elliptic topological vertex}
\dedicatory{To the memory of Professor Vladimir Rittenberg}
	
\author[]{Omar Foda      \!$^{{\scriptstyle {\, 1            }}}$ and 
          Rui-Dong Zhu   \!$^{{\scriptstyle {\, 2            }}}$
          }
\address{
$^{{\scriptstyle 1}}$
School of Mathematics and Statistics, 
The University of Melbourne, Parkville, Victoria 3010, Australia
\newline
$^{{\scriptstyle 2}}$
Department of Physics, The University of Tokyo, 
Bunkyo-ku, Tokyo, Japan
}

\email{omar.foda@unimelb.edu.au,
nick\_zrd@hep-th.phys.s.u-tokyo.ac.jp}

	
\begin{abstract}
We derive a one-parameter deformation of the refined topological vertex that,
when used to compute non-periodic web diagrams, reproduces the six-dimensional 
topological string partition functions that are computed using the refined vertex 
and periodic web diagrams.
\end{abstract}
	
\maketitle

\section{Introduction}
\label{section.01}
\textit{We motivate the study of topological vertices from a 2D conformal field theory 
point of view.}

\subsection{A web of relations}
The relation of 2D conformal field theories, which describe critical surface phenomena, and 
exact solutions in 2D statistical mechanical models, which describe off-critical surface
phenomena, is well-understood since the 1970's. More recently, 4D, 5D and 6D instanton and 
topological string partition functions, as well as other topics in modern mathematical physics, 
were related to 2D conformal field theories in terms of \textit{dualities}, and one can study 
any of these topics from the viewpoint of any of the others
\footnote{\,
For a review of recent developments, see \cite{teschner.review, pestun.zabzine.review}
}. In the following, we motivate the present work from the viewpoint of 2D conformal field 
theory.

\subsection{From 2D correlation functions to plane partitions} 
In 1984, Belavin, Polyakov and Zamolodchikov showed that correlation functions in 2D conformal 
field theories split into sums of products of structure constants, chiral conformal blocks, 
and anti-chiral conformal blocks. In 2009, Alday, Gaiotto and Tachikawa conjectured 
\cite{alday.gaiotto.tachikawa}, and Alba, Fateev, Litvinov and Tarnopolskiy proved 
\cite{alba.fateev.litvinov.tarnopolskiy}, that in the presence of an extra Heisenberg algebra, 
a 2D conformal block splits into products of 4D Nekrasov partition functions
\footnote{\,
Alday \textit{et al.} also conjecture 4D interpretations of the structure constants as well as 
other aspects of 2D correlation functions, but in the present work, we focus on the conformal 
blocks.
}, which are limits of 5D instanton partition functions. These 5D instanton partition functions 
split into products of topological vertices that are also 5D partition functions 
\footnote{\, 
Topological vertices are 5D partition functions in the sense that they depend $R$, the radius 
of the $M$-theory circle. Gluing topological vertices leads to $R$-deformed 2D conformal blocks. 
In 2D terms, $R$ is an off-critical deformation parameter, and the critical 2D conformal blocks 
are obtained in the $R \rightarrow 0$ limit. 
}. 
Since a topological vertex has a combinatorial interpretation as a generating function of 
weighted plane partitions that satisfy specific boundary conditions 
\cite{okounkov.reshetikhin, okounkov.reshetikhin.vafa}, 
the (difficult) analytic problem of computing correlation functions in 2D conformal field 
theory is recast as a (hopefully) simpler exercise in algebraic combinatorics.

\subsection{The algebraic combinatoric point of view}
Viewing the 2D correlation functions in terms of algebraic combinatorial objects, which 
in this case are plane partitions with specific weights and specific boundary conditions, 
allows one to study more general classes of them. One way to do that is to change the weights 
while maintaining computability. All known topological vertices, starting from the original 
vertex $\cO_{\, Y_1 Y_2 Y_3} \ll x \rr$, 
which depends on three Young diagrams $Y_1, Y_2$ and $Y_3$, and a single parameter $x$
\footnote{\,
In this work, as in \cite{foda.wu.02}, we use $x$ and $y$ for the parameters of the refined 
vertex, 
$x = \exp \ll - R \epsilon_1 \rr$ and
$y = \exp \ll   R \epsilon_2 \rr$, 
where $\epsilon_1$ and $\epsilon_2$ are Nekrasov's deformation parameters \cite{nekrasov}, 
and $R$ is the radius of the M-theory circle. We reserve $q$ and $t$ for the Macdonald 
deformation 
parameters \cite{macdonald.book}.
}, 
and leads to conformal blocks in conformal field theories with integral central charges
\cite{aganagic.klemm.marino.vafa}, 
to the refined vertex $\cR_{\, Y_1 Y_2 Y_3} \ll x, y \rr$, 
which depends on an additional refinement parameter $y$, 
and leads to conformal blocks in conformal field theories with generic central charges
\cite{awata.kanno.01, awata.kanno.02, iqbal.kozcaz.vafa}, 
to the Macdonald vertex $\cM_{\, Y_1 Y_2 Y_3} \ll x, y \, | \, q, t \rr$ 
which depends on two additional Macdonald parameters $q$ and $t$ \cite{foda.wu.02}, 
and leads to conformal blocks in the presence of vertex-operator condensates 
\cite{foda.manabe}, are generating functions of plane partitions that are given 
different weights.

\subsection{The present work}
Following Saito's construction of an elliptic version of Ding-Iohara-Miki algebra 
\cite{ding.iohara, miki}, using \textit{two} commuting Heisenberg algebras, one 
deformed by $q$, and the other by $1/q$
\cite{saito.01, saito.02, saito.03}, we construct an \textit{elliptic} 
vertex 
$\cE_{\, \bY_1 \bY_2 Y_3}$ $\ll \, x, \, y \, | \, \bfq \, \rr$, where 
$\bY_1 = \ll Y_{\, 1 \, A}, Y_{\, 1 \, B} \rr$ and 
$\bY_2 = \ll Y_{\, 2 \, A}, Y_{\, 2 \, B} \rr$ are pairs of Young diagrams, 
$Y_3$ is a single Young diagram, and
$\bfq  = \ll q, 1/q \rr$, where $q$ is a deformation parameter. 

\subsubsection{Two components} $\cE$ is a product of two components,

\begin{equation}  
\cE_{\, \bY_{1 }  \bY_{2 }  Y_3}  \ll \, x,   \, y   \, | \, \bfq \rr = 
\cM_{\,   Y_{\, 1 \, A}    Y_{\, 2 \, A}  Y_3}  \ll \, x,   \,   y \, | \,    q \, \rr \, 
\cM_{\,   Y_{\, 1 \, B}    Y_{\, 2 \, B}  Y_3}  \ll \, 1/x, \, 1/y \, | \,  1/q \, \rr, 
\end{equation} 

where $\cM_{\, Y_{\, 1 \, A} Y_{\, 2 \, A} Y_3} \ll \, x, \, y \, | \, q \, \rr$ is 
a Macdonald vertex with Macdonald parameters $q \neq 1$, and $t = 0$, and refinement 
parameters $x$ and $y$, and 
$\cM_{\, Y_{\, 1 \, B} Y_{\, 2 \, B} Y_3} \ll \, 1/x, \, 1/y \, | \, 1/q \, \rr$ is 
a Macdonald vertex with Macdonald parameters $1/q \neq 1$, and $t = 0$, 
and refinement parameters $1/x$ and $1/y$.
The Young diagrams $Y_{\, 1 \, A}$ and $Y_{\, 1 \, B}$ that label the initial non-preferred 
legs of the component Macdonald vertices are independent, 
the Young diagrams $Y_{\, 2 \, A}$ and $Y_{\, 2 \, B}$ that label the final non-preferred 
legs are also independent, but the same Young diagram $Y_3$ labels the (common) preferred 
leg of both component vertices
\footnote{\,
These terms will be defined when we construct the elliptic vertex explicitly in section 
\ref{section.09}.
}.
The original Macdonald vertex depends on two Macdonald parameters $q$ and $t$ and has 
basically the same structure as the refined topological vertex, but the Schur 
functions replaced by Macdonald functions. 
The component Macdonald vertices depend on a single Macdonald parameter $q$ or $1/q$, 
and the second Macdonald parameter $t=0$. In this case, the Macdonald functions are 
one-parameter deformations of the Schur functions called
$q$-Whittaker functions 
\cite{
gerasimov.lebedev.oblezin.01, 
gerasimov.lebedev.oblezin.02,
gerasimov.lebedev.oblezin.03,
borodin.corwin,
borodin.petrov, 
borodin.wheeler}.

\subsubsection{The refined vertex limit} In the limit $q \rightarrow 0$, 

\begin{multline}
\cM_{\,   Y_{\, 1 \, A}    Y_{\, 2 \, A}  Y_3}  \ll \, x,   \, y   \, | \,    q \, \rr \rightarrow
\cR_{\,   Y_{\, 1 \, A}    Y_{\, 2 \, A}  Y_3}  \ll \, x,   \, y \, \rr, 
\quad
\cM_{\,   Y_{\, 1 \, B}    Y_{\, 2 \, B}  Y_3}  \ll \, 1/x, \, 1/y \, | \,  1/q \, \rr \rightarrow 1, 
\\
\cE_{\, \bY_{1 }  \bY_{2 }  Y_3}  \ll \, x,   \, y   \, | \, \bfq \rr
\rightarrow
\cR_{\,   Y_{\, 1 \, A}    Y_{\, 2 \, A}  Y_3}  \ll \, x,   \, y \, \rr, 
\end{multline}

where $\cR_{\,   Y_{\, 1 \, A}    Y_{\, 2 \, A}  Y_3}  \ll \, x,   \, y \, \rr$ is a refined vertex
\footnote{\,
In this work, we always refer to the formulation of the refined vertex in \cite{iqbal.kozcaz.vafa}
}. 
In the limit $q \rightarrow \infty$,

\begin{multline}
\cM_{\,   Y_{\, 1 \, A}    Y_{\, 2 \, A}  Y_3}  \ll \, x,   \, y   \, | \,    q \, \rr \rightarrow 1,
\quad 
\cM_{\,   Y_{\, 1 \, B}    Y_{\, 2 \, B}  Y_3}  \ll \, 1/x, \, 1/y \, | \,  1/q \, \rr \rightarrow 
\cR_{\,   Y_{\, 1 \, B}    Y_{\, 2 \, B}  Y_3}  \ll \, 1/x,   \, 1/y \, \rr, 
\\
\cE_{\, \bY_{1 }  \bY_{2 }  Y_3}  \ll \, x,   \, y   \, | \, \bfq \rr \rightarrow 
\cR_{\,   Y_{\, 1 \, B}    Y_{\, 2 \, B}  Y_3}  \ll \, 1/x,   \, 1/y \, \rr
\end{multline}

In this sense, the elliptic vertex 
$\cE_{\, \bY_{1 }  \bY_{2 }  Y_3}  \ll \, x,   \, y   \, | \, \bfq \rr$ is a one-parameter deformation 
of the refined vertex that interpolates
$\cR_{\,   Y_{\, 1 \, A}    Y_{\, 2 \, A}  Y_3}  \ll \, x,   \, y \, \rr$ and 
$\cR_{\,   Y_{\, 1 \, B}    Y_{\, 2 \, B}  Y_3}  \ll \, 1/x,   \, 1/y \, \rr$, which are equivalent in 
the sense that they lead to the same 4D and 5D instanton partition functions and the same 2D conformal 
blocks, up an overall normalization.

\subsubsection{Twisted vertices}
We also introduce \textit{twisted} version of $\cE$ that we call $\cE^{\, \star}$, that 
depend on \textit{twisted} $q$-Whittaker functions that we define. 

\subsubsection{6D instanton partition functions from non-periodic web diagrams} 
Starting from a non-periodic web diagram, constructed by gluing refined topological vertices, such that 
the corresponding partition function is a 5D instanton partition function, and replacing the refined 
vertices by $\cE$ and $\cE^{\, \star}$ alternately, we reproduce the 6D version of the 5D instanton 
partition function that we start with, while keeping the connectivity of the web diagram intact. 
These 6D instanton partition functions are the same as those obtained by taking traces, that is, 
by identifying opposite external legs and summing over the intermediate states to form periodic web 
diagrams, then computing their partition functions, as first proposed in the work of Hollowood, Iqbal 
and Vafa \cite{hollowood.iqbal.vafa}
\footnote{\,
The literature on 6D instanton partition functions and related topics has grown rapidly over the past few years. The present work was motivated by the recent works of Iqbal, Kozcaz and Yau 
\cite{iqbal.kozcaz.yau} and Nieri \cite{nieri}.
}. 

\subsection{Two routes} We came to the elliptic vertex $\cE$ \textit{via} two different 
routes.

\subsubsection{The Iqbal-Kozcaz-Vafa refined vertex route}
\label{route.01}
In \cite{foda.wu.02}, the first co-author, together with Jian-Feng Wu, proposed a Mac\-donald-type
deformation of the refined vertex of Iqbal, Kozcaz and Vafa \cite{iqbal.kozcaz.vafa}, in terms 
of the Macdonald parameters $q$ and $t$, and noted its connection to Ding-Iohara-Miki algebra
\cite{ding.iohara, miki, feigin.hashizume.hoshino.shiraishi.yanagida}. In \cite{foda.wu.03}, 
an elliptic extension of the Macdonald vertex of \cite{foda.wu.02} based on Saito's elliptic 
extension of the Ding-Iohara-Miki algebra was obtained. This extension introduced a parameter 
$p$, and represents the initial and final states of the new vertex in terms of pairs of
$p$-deformed Macdonald functions
\footnote{\,
Saito's elliptic extension of the Ding-Iohara-Miki algebras makes use of two Heisenberg 
algebras, and the corresponding states are naturally represented in terms of pairs of 
symmetric functions \cite{wu.private.communication}.
}. 
The properties of these $p$-deformed Macdonald functions were not completely well-understood, 
and to work with them, one had to conjecture that they satisfied suitable Cauchy-type 
identities. However, setting $q=t$, the $p$-deformed Macdonald functions reduced to 
$p$-deformed Schur functions, whose properties were still not completely well-understood, 
but appeared to be more amenable to analysis \cite{foda.unpublished}. 

\subsubsection{The Awata-Feigin-Shiraishi refined vertex route}
\label{route.02}
In \cite{zhu.01}, the second co-author proposed a Saito-type elliptic 
extension of the refined vertex of Awata, Feigin and Shiraishi 
\cite{awata.feigin.shiraishi}, and in \cite{zhu.02}, he obtained another version in a form 
related to that of Iqbal, Kozcaz and Vafa \cite{zhu.02}, where the initial and final states 
are also labelled by pairs of $p$-deformed Schur functions. This elliptic vertex is exactly 
that of \cite{foda.wu.03}, restricted to the $p$-deformed Schur functions of \cite{foda.unpublished}.

\subsubsection{$q$-Whittaker functions} 
The starting point of the present work is the observation that the $p$-deformed Schur 
functions of sections \ref{route.01} and \ref{route.02} are $q$-Whittaker functions 
obtained from Macdonald functions in the limit $t \rightarrow 0, q \neq 0$
\footnote{\,
The $q$-Whittaker functions are dual to the Hall-Littlewood functions in the sense that 
the former are obtained in the limit $t \rightarrow 0, q \neq 0$ of Macdonald functions, 
while the latter are obtained in the limit 
$q \rightarrow 0, t \neq 0$.
},
where the Macdonald parameter $q$ plays the role of Saito's elliptic deformation parameter 
$p$
\footnote{\,
In Saito's work \cite{saito.01, saito.02, saito.03}, and in the present work, 
whenever the parameters $p$ and $q$ are both non-zero, they appear on equal footing.
}. 
This observation allows us to use the tools of Macdonald functions, in the limit 
$t \rightarrow 0, q \neq 0$, to put our derivations on a solid footing. 

\subsection{Outline of contents}
In section \ref{section.02}, we recall basic facts related to $q$-Whittaker functions and 
their Cauchy identities, then in
in section \ref{section.03}, we introduce an involution that we use to define twisted 
$q$-Whittaker function and their Cauchy identities.
In section \ref{section.04}, we recall Saito's pair of $pqt$-Heisenberg algebras and pair 
of $pqt$-vertex operators.
In each pair, one component depends on $p, q$ and $t$, and the other on $1/p, 1/q$ and 
$1/t$.
We show that Macdonald's parameter $q$ and Saito's parameter $p$ appear in all expressions on 
equal footing, so that setting $q=t$, or $p=t$, all dependence on the equated parameters
disappears, and the remaining parameter can be identified with the parameter that deforms Schur
functions to $q$-Whittaker functions. 
In section \ref{section.05}, we take the $p \rightarrow t$ limit of Saito's $pqt$-Heisenberg 
algebras and $pqt$-vertex operators, to obtain a pair of $q$-deformed Heisenberg algebras, and 
a pair of $q$-deformed vertex operators, such that in each pair, one component depends on $q$ 
and the other depends on $1/q$.
In section \ref{section.06}, we recall the Heisenberg/power sum correspondence which allows us 
to derive useful operator-valued identities for one of the Heisenberg algebras of section 
\ref{section.05}, 
then in section \ref{section.07}, we do the same for the second Heisenberg algebra.
In section \ref{section.08}, we define pairs of $q$-Whittaker functions and derive their Cauchy 
identities.
In section \ref{section.09}, we construct the elliptic vertex, and 
in section \ref{section.10}, we show that gluing copies of this vertex produces an elliptic 
version of the strip partition function \cite{iqbal.kashani-poor}, so that gluing copies of 
the latter produces the 6D instanton partition functions of \cite{hollowood.iqbal.vafa}.
Section \ref{section.11} includes a number of comments.

\subsection{Notations and other conventions}

\subsubsection{Sets}
$\bi$, and similarly $\bj$, is the set of non-zero natural numbers $\ll 1, 2, \cdots \rr$,
$\bx = \ll x_1, x_2, \cdots \rr$ and 
$\by = \ll y_1, y_2, \cdots \rr$ are sets of (possibly infinitely-many) variables, 
$\ba_{-} = \ll a_{-1}, a_{-2}, \cdots \rr$ and   
$\ba_{+} = \ll a_{ 1}, a_{ 2}, \cdots \rr$ are 
the free-boson creation and annihilation mode operators.

\subsubsection{Pairs of Young diagrams and pairs of variables}
$\bY = \ll Y_1, Y_2 \rr$ is a pair of Young diagrams,
$\pmb{\emptyset} = \ll \emptyset, \emptyset \rr$ is a pair of empty Young diagrams, and 
$\bfq$ is the pair 
$\ll q, 1/q \rr$. 

\subsubsection{Number of elements in sets and number of cells in Young diagrams}
$| \, \bx \, |, \, | \, \by \, |, \, \cdots,$ are the numbers of elements in the sets 
$\bx, \, \by, \, \cdots,$ and $| \, \bY \, | = | \, Y_1 \, | + | \, Y_2 \, |$, where 
$| \, Y_i \, |$ is the number of cells in the Young diagram $Y_i$.

\subsubsection{Primed variables and transpose Young diagrams}
To simply the notation, we use the primed variables 
$x^{\, \prime}, y^{\, \prime}, p^{\, \prime}, q^{\, \prime}, t^{\, \prime}, \cdots$,  
for the inverse variables
$1/x, 1/y, 1/p, 1/q, 1/t, \cdots$, 
and the primed set variable 
$\, \bx^{\, \prime} = 
\ll x_1^{\, \prime}, x_2^{\, \prime}, \cdots \rr$ 
for the set of inverse variables
$\ll 1/x_1,    1/x_2,     \cdots \rr$.
The Young diagram $Y^{\, \prime}$ is the transpose of the Young diagram $Y$.  
We use $W^{\, \prime}_q$ for the dual $q$-Whittaker symmetric function as in 
section \ref{macdonald.q.whittaker}. 

\subsubsection{Macdonald and $q$-Whittaker symmetric functions}
\label{macdonald.q.whittaker}
We use $P_{\, Y} \ll \bx \rr$ and $Q_{\, Y} \ll \bx \rr$ for the Macdonald and dual Macdonald 
symmetric functions. 
Each of these functions is labelled by a Young diagram $Y$, depends on two parameters $q$ and $t$, 
and is symmetric in a (possibly infinite) set of variables $\bx = \ll x_1, x_2, \cdots \rr$. We use 
$W_{\, q \,  Y} \ll \bx \rr$ and $W^{\, \prime}_{\, q \,  Y} \ll \bx \rr$ for the $q$-Whittaker and 
dual $q$-Whittaker symmetric functions. Each of these functions is labelled by a Young diagram $Y$, 
a parameter $q$, and is symmetric in a (possibly infinite) set of variables $\bx = \ll x_1, x_2, 
\cdots \rr$
\footnote{\, 
We show the dependence on the variable $q$ explicitly because this will often be a dependence on 
$q^{\prime} = 1/q$.
}.

\subsubsection{Pairs of $q$-Whittaker symmetric functions}
We use 
$\bW^{         }_{\, \bfq  \, \bY} \ll \bx \rr$ and 
$\bW^{\, \prime}_{\, \bfq  \, \bY} \ll \bx \rr$,
$\bfq = \ll q, 1/q \rr$, 
$\bY  = \ll Y_1, Y_2 \rr$, for 
a pair of      $q$-Whittaker and 
a pair of dual $q$-Whittaker symmetric functions. 
The first symmetric function in a pair depends 
on a parameter $q$, a Young diagram $Y_1$, and 
is symmetric in a (possibly infinite) set of variables $\bx = \ll x_1, x_2, \cdots \rr$, while
the second symmetric function in the pair depends 
on a parameter $1/q$, a Young diagram $Y_2$, and 
is symmetric in a (possibly infinite) set of variables $\bx^{\, \prime} = \ll x_1^{\, \prime}, x_2^{\, \prime}, \cdots \rr$. 

\subsubsection{Parameters}
Our refinement parameters $\ll x, y\rr$ are the parameters $\ll q, t \rr$ in \cite{iqbal.kozcaz.vafa}
\footnote{\,
More precisely, our $x$ is $t$, and our $y$ is $q$ in \cite{iqbal.kozcaz.vafa}
}.
Our $q$-Whittaker deformation parameter (which will be called either $q$ or $1/q$) is the first Macdonald parameter 
$q$, while the second Macdonald parameter 
$t=0$. 

\subsubsection{Exponentiated sequences}
\label{sequences}
Given a Young diagram $Y$ that consists of an infinite sequence of rows $Y = \ll y_1, y_2, 
\cdots \rr$, such that only finitely-many rows have non-zero length, together with an infinite 
sequence of integers $\bi = \ll 1, 2, \cdots \rr$, and two variables $u$ and $v$, we define 
the exponentiated sequences $u^{\, \bi}$, $v^{\, \pm Y}$, $\cdots$, as, 

\begin{equation}
u^{\, \bi} = \ll u, u^2, \cdots \rr,
\quad
u^{\, \bi - 1} = \ll 1, u,   \cdots \rr,
\quad  
v^{\, \pm Y}    = \ll v^{\pm y_1}, v^{\pm y_2}, \cdots \rr,
\quad
\cdots, 
\end{equation}

and the products of exponentiated sequences $u^{\, \bi}\, v^{\, \pm Y}$, 
$u^{\, \bi - 1}\, v^{\, \pm Y}$, $\cdots$, as, 

\begin{equation}
u^{\, \bi}\, v^{\, \pm Y}  = \ll u  \, v^{\, \pm y_1}, u^2\, v^{\, \pm y_2} \cdots \rr, 
\quad
u^{\, \bi - 1}\, v^{\, \pm Y}  = \ll v^{\, \pm y_1}, u\, v^{\, \pm y_2} \cdots \rr, 
\quad
\cdots 
\end{equation}

\subsubsection{More on sequences}
\label{more.on.sequences}
Let 
$\bx = \ll x_1, \cdots, x_m \rr$ be a set of $m$ variables, and 
$Y =   \ll y_1, \cdots, y_n \rr$ be a Young diagram that consists of $n$ non-zero 
parts, such that $n \leq m$. The notation, 
	
\begin{equation}
\bx_{\bi}^Y = \ll x_{\iota_1}^{y_1}, \, \cdots, \, x_{\iota_n}^{y_n} \rr, 
\label{sum}
\end{equation}
	
where $\bi = \ll \iota_1, \cdots, \iota_n \rr$, is defined as follows. 
\1 Consider the set of $m$ integers, ${\pmb m} = \ll 1, \cdots, m \rr$, 
for example, ${\pmb m} = \ll 1, 2, 3, 4 \rr$, 
\2 Choose a subset of $n$ integers ${\pmb n} \subseteq {\pmb m}$, 
for example, ${\pmb n} = \ll 1, 2, 4 \rr$, 
\3 Consider a specific permutation $\bi$ of ${\pmb n}$, for example, $\bi = \ll 2, 4, 1 \rr$.
\4 The set on the right hand side of  Equation \ref{sum} is obtained by starting with the set 
$\ll x_{\iota_1}, \, \cdots, \, x_{\iota_n} \rr$, 
and raising its elements sequentially to the powers
$\ll y_1,         \, \cdots, \, y_n         \rr$
\footnote{\, 
In applications of this notation, for example to the definition of the monomial symmetric 
functions, one sums over all permutations $\bi$ of all possible distinct subsets ${\pmb n}$
of the same cardinality.
}. 

\subsubsection{Products on sequences}
\label{products.on.sequences}
We will use the notation, 

\begin{multline}
\frac{1}
     {1\, -\, \bx \, \by \, q^{\, n}}
= 
    \prod_{i,\, j = \, 1}^\infty
\ll 
\frac{1}
     {1\, -\, x_i\, y_j \, q^{\, n}}
\rr,
\quad 
1\, + \, \bx \, \by \, q^{\, n}
= 
    \prod_{i,\, j = \, 1}^\infty
\ll 
1\, + \, x_i\, y_j \, q^{\, n}
\rr,
\\
\phi_{\, q \, \pm} \ll \bx \rr = 
\prod_{i=1}^{\, \infty} \phi_{\, q \, \pm} \ll x_{\, i} \rr,
\quad 
\Gamma_{\, q \, a \, \pm} \ll \bx \rr 
=
\prod_{i \, =\, 1}^\infty \Gamma_{\, q \, a \, \pm} \ll x_i \rr, 
\quad
\Gamma^{\, \pm}_{\, q \, b \, \pm} \ll \bx \rr 
=
\prod_{i \, =\, 1}^\infty 
\Gamma^{\, \pm}_{\, q \, b \, \pm} \ll x_i \rr,
\label{abbreviation}
\end{multline}

where 
$\phi_{\, q \, \pm} \ll x_{\, i}^{\, \prime} \rr$, \textit{etc.} 
are two-boson vertex operators, and 
$\Gamma_{\, q \, a \, -} \ll x_i \rr$, and
$\Gamma^{\, \pm}_{\, q \, b \, -} \ll x_i \rr$ 
are one-boson vertex operators, to be defined in section \ref{section.04}.

\subsubsection{Products on almost theta functions}
We will also use,

\begin{equation}
\theta_{\, q} \ll \bx, \, \by \rr
=
\prod_{i,\, j = 1}^\infty
\theta_{\, q} \ll x_i \, y_j \rr, 
\quad
\theta_{\, q} \ll x_i \, y_j \rr
=
\prod_{n = 0}^\infty 
\ll 1\, -\, x_i \, y_j \, q^{\, n  } \rr
\ll 1\, -\, x^{\, \prime}_i \, y^{\, \prime}_j \, q^{ \, n+1  } \rr, 
\label{abbreviation.02}
\end{equation}

that is, $\theta_{\, q} \ll x \rr$ is a Jacobi theta function 
$\Theta \ll x \, | \,  q \rr$, up to an $x$-independent factor, 

\begin{multline}
\theta_{\, q} \ll x \rr = \Theta \ll x \, | \, q \rr / \ll q \, | \, q \rr, 
\\ 
\Theta \ll x; q \rr =  
\prod_{n = 0}^\infty 
\ll 1\, -\, q^{\, n + 1 } \rr
\ll 1\, -\, x \, q^{\, n  } \rr
\ll 1\, -\, x^{\, \prime}_i\, q^{ \, n + 1  } \rr,   
\quad  
\ll q \, | \, q \rr = 
\prod_{n = 1}^\infty
\ll 1\, - \, q^{\, n} \rr 
\end{multline}

\section{The $q$-Whittaker functions}
\label{section.02}
\textit{Starting from the properties of the Macdonald functions, which depend on $q$ and $t$, we take the limit $t \rightarrow 0$ to obtain the corresponding properties of the
$q$-Whittaker functions, which depend on $q$ only.
}
\smallskip 

\subsection{The monomial symmetric functions}
$m_{\, Y} \ll \bx \rr$, where $\bx = \ll x_1, x_2, \cdots \rr$, indexed by a Young diagram $Y$, is
\footnote{\, 
Ch. I, p. 18, Equation 2.1, in \cite{macdonald.book}
}, 
	
\begin{equation}
m_{\, Y} \ll \bx \rr = \sum_{\bi} x_{\bi}^Y,  
\end{equation}

where the sum runs over all  distinct permutations of the set $\bi$, which is defined as in section \ref{more.on.sequences}. For example, 

\begin{equation}
m_\emptyset \ll \bx \rr = 1,            \, \, 	
m_1         \ll \bx \rr = \sum_i x_i,   \, \, 
m_2         \ll \bx \rr = \sum_i x_i^2, \, \,  
\cdots, 
\, \, 
m_{443}     \ll x_1, \cdots, x_5 \rr = \sum_{\bi} x_i^4 \, x_j^4 \, x_k^3, 	
\end{equation}

where the sum in the last example is over all distinct permutations $\bi$, of all distinct 
subsets ${\pmb m} \subseteq {\pmb n} = \ll 1, \cdots, 5 \rr$, such that the cardinality 
$| \, {\pmb m} \, | = 3$, and $i \neq j \neq k \in {\pmb n}$, as defined in section \ref{more.on.sequences}. 

\subsection{Power-sum symmetric functions}
$p_n \ll \bx \rr$,
where $\bx = \ll x_1, x_2, \cdots \rr$, 
indexed by an integer $n \in \ll 0, 1, \cdots \rr$,
is
\footnote{\,
Ch. I, p. 23, in \cite{macdonald.book}
}, 
	
\begin{equation}
p_0 \ll \bx \rr = 1,
\quad  
p_n \ll \bx \rr = \sum_i x_i^{\, n} = m_n \ll \bx \rr,
\quad 
n = 1, 2, \cdots,  
\end{equation}
	
and $p_{\, Y} \ll \bx \rr$, indexed by a Young diagram $Y = \ll y_1, y_2, \cdots \rr$, is
\footnote{\,
Ch. I, p. 24, in \cite{macdonald.book}
}, 
	
\begin{equation}
    p_{\, Y}     \ll \bx \rr\, =\, 
    p_{\, Y_1} \ll \bx \rr\, 
    p_{\, Y_2} \ll \bx \rr\, \cdots
\end{equation}

\subsection{$q$-Whittaker functions as $t \rightarrow 0$ limits of Macdonald 
functions} 
Consider the ring of symmetric functions in the variables $\bx = \ll x_1, x_2, \cdots \rr$, 
with coefficients in the field of rational functions in two variables $\ll q, t\rr$. 
In this ring, 
the      Macdonald functions $P_{\, Y} \ll \bx \rr$, each      labelled by a Young diagram $Y$, and 
the dual Macdonald functions $Q_{\, Y} \ll \bx \rr$, each also labelled by a Young diagram $Y$, 
form two orthogonal bases.
In the limit $t \rightarrow 0$, the coefficients of the ring of symmetric functions are in 
the field of rational functions in a single variable $q$, 
the      Macdonald functions $P_{\, Y} \ll \bx \rr$ reduce to the      $q$-Whittaker functions, 
which we denote by $W_{\, q \, Y} \ll \bx \rr$, and 
the dual Macdonald functions $Q_{\, Y} \ll \bx \rr$ reduce to the dual $q$-Whittaker functions,  
which we denote by $W^{\, \prime}_{\, q \, Y} \ll \bx \rr$.
These functions were introduced by Gerasimov, Lebedev and Oblezin 
\cite{gerasimov.lebedev.oblezin.01, gerasimov.lebedev.oblezin.02, gerasimov.lebedev.oblezin.03}, 
and further studied in  
\cite{borodin.corwin, borodin.petrov, borodin.wheeler}. 
In the rest of this section, we deduce the properties of and relations satisfied by 
$W_{\, q \, Y} \ll \bx \rr$ and 
$W^{\, \prime}_{\, q \, Y} \ll \bx \rr$
by taking the $t \rightarrow 0$ limit of the corresponding properties and relations 
satisfied by $P_{\, Y} \ll \bx \rr$ and $Q_{\, Y} \ll \bx \rr$. 

\subsection{The $q$-inner product of the power-sum symmetric functions}
\label{power.sum.inner.product.macdonald.basis}
From the orthogonality of 
the power-sum symmetric functions in the ring of symmetric functions with 
coefficients in the field of rational functions in $q$ and $t$
\footnote{\,
Ch. VI, p. 225, Equation 4.11, in \cite{macdonald.book}
}, 
the power-sum symmetric functions in the ring of symmetric functions with 
coefficients in the field of rational functions in $q$ 
are orthogonal with respect to the $q$-inner product,  
	
\begin{equation}
\langle\, p_{\, Y_1} \ll \bx \rr\, |\, p_{\, Y_2} \ll \bx \rr \rangle_{\, q} = 
z_{\, q \, Y_1}\, \delta_{Y_1 Y_2},
\quad
z_{\, q \, Y} 
=
\ll
1^{n_1} \ll n_1 ! \rr
2^{n_2} \ll n_2 ! \rr 
\cdots
\rr
\prod_{i=1}^{y_1^{\, \prime}} 
\ll
1 - q^{\, y_i}
\rr,  
\label{young.diagram.power.sum.inner.product.a}
\end{equation} 

where $n_{\, r}$ is the number of rows of length $r$ in $Y$, and $y^{\, \prime}_1$
is the length of the first row in $Y^{\, \prime}$, that is, the number of non-zero rows in $Y$. This inner product can be understood 
as follows
\footnote{\, 
Ch. I, p. 75--76, in \cite{macdonald.book}
}. For every power-sum symmetric 
function $p_{\, Y} \ll \bx \rr$, there is a differential operator 
$D_{\, Y} \ll \bx \rr$ in $\bx = \ll x_1, x_2, \cdots \rr$, such that acting with 
$D_{\, Y} \ll \bx \rr$ on $p_{\, Y} \ll \bx \rr$, then setting 
$x_1 = x_2 = \cdots = 0$, one obtains the right hand side of the first of Equations 
\ref{young.diagram.power.sum.inner.product.a}.

\subsection{A $q$-identity}
The power-sum symmetric functions $p_n \ll \bx \rr$ satisfy the $q$-identity, 

\begin{equation}
\exp 
\ll
\sum_{n=1}^{\infty} \, 
\frac{1}{n} 
\frac{1}{\ll 1 - q^{\, n} \rr} \, 
p_n \ll \bx \rr 
p_n \ll \by \rr
\rr
=
\prod_{n         = \, 0}^\infty
\ll 
\frac{
1
}{
1\, -\, \bx \, \by \, q^{\, n}}
\rr, 
\label{an.exponential.is.a.product}
\end{equation}

which follows from expanding the exponent on the left hand side, then using, 

\begin{equation}
\exp \ll - \sum_{n = 1}^{\infty} \frac{x^{\, n}}{n} \rr = 
\exp \ll \log \ll 1 - x \rr \rr =
1 - x,       
\end{equation}

to resum the result of the expansion in the form of the right hand side. 

\subsection{The $q$-Whittaker function}
From the definition of the Macdonald function $P_{\, Y} \ll \, \bx \rr$
\footnote{\,
Ch. VI, p. 322, in \cite{macdonald.book}
}, 
we obtain the $q$-Whittaker function 
$W_{\, q \, Y} \ll\, \bx\, \rr$, $Y = \ll y_1, y_2, \cdots \rr$, as the unique symmetric function in 
$\bx = \ll x_1, x_2, \cdots \rr$, $| \, \bx \, | \geq y_1^{\, \prime}$, that satisfies two 
properties.

\subsubsection{The expansion in terms of monomial symmetric functions} 
 
\begin{equation}
W_{\, q \,  Y_1} \ll\, \bx\, \rr =
m_{\, Y_1} \ll \bx \rr \, + \,  
\sum_{Y_1 \succ Y_2} \, u_{\, q \, Y_1 \, Y_2} \, m_{\, Y_2} \ll \bx \rr, 
\end{equation}
	
where $m_{\, Y} \ll \bx \rr$ is the monomial symmetric function in $\bx$ labelled by $Y$, 
$Y_1 \succ Y_2$ indicates that $Y_1$ dominates $Y_2$ in the natural partial ordering of Young diagrams
\footnote{\,
Ch. I, p. 7, in \cite{macdonald.book}
}, 
and the coefficients $u_{\, q \, Y_1 \, Y_2}$ are rational functions in $q$. 

\subsubsection{The orthogonality relation}  

\begin{equation}
\langle 
W_{\, q \,  Y_1} \ll\, \bx\, \rr\, |\, 
W_{\, q \,  Y_2} \ll\, \bx\, \rr 
\rangle = 0, 
\quad
\textit{for}
\quad
Y_1 \neq \, Y_2
\label{macdonald.orthogonality}
\end{equation}

\subsection{The dual $q$-Whittaker function} 
From the definition of the dual Macdonald function $Q_{\, Y} \ll \, \bx \, \rr$
\footnote{\,
Ch. VI, p. 322, in \cite{macdonald.book}
},
the dual $q$-Whittaker function $W^{\, \prime}_{\, q \, Y} \ll\, \bx\, \rr$ is 
defined in terms of $W_{\, q \, Y} \ll \, \bx\, \rr$ as
\footnote{\,
Ch. VI, p. 323, Equation 4.12, and 
        p. 339, Equation 6.19, in \cite{macdonald.book}
},
	
\begin{equation}
W^{\, \prime}_{\, q \, Y} \ll\, \bx\, \rr  = 
b_{\, Y}   
\, 
W_{\, q \, Y} \ll\, \bx\, \rr, 
\quad  
b_{\, Y} = 
\prod^{\, \prime}_{\wsq \, \in\, Y}
\ll 
\frac{
1
}{
1\, -\,q^{\, A_{\wsq \, Y}^+} 
}
\rr, 
\label{dual.macdonald}
\end{equation}

where the prime on the product indicates that the product is restricted to cells $\wsq \, \in\, Y$ 
with leg-length $L_{\, \wsq} = 0$. This is obtained as follows. For $q \neq 0$, the product on the 
right hand side of Equation \ref{dual.macdonald} is,

\begin{equation}
b^{\, q \, t}_{\, Y} = 
\prod_{\wsq \, \in\, Y}
\ll 
\frac{
1\, -\,q^{\, A_{\wsq \, Y}} \, t^{\, L_{\wsq \, Y}^+} 
}{
1\, -\,q^{\, A_{\wsq \, Y}^+} \, t^{\, L_{\wsq \, Y}} 
}
\rr    
\end{equation}

In the limit $t \rightarrow 0$, 
the numerator   $1\, -\,q^{\, A_{\wsq \, Y}} \, t^{\, L_{\wsq \, Y}^+} \rightarrow 1$,
for all $\wsq \, \in\, Y$, while
the denominator $1\, -\,q^{\, A_{\wsq \, Y}^+} \, t^{\, L_{\wsq \, Y}} \rightarrow 1$,  
for all $\wsq \, \in\, Y$ such that $L_{\wsq \, Y} \geq 1$, and the corresponding 
factors trivialize to 1. All non-trivial contributions are due to 
 $\wsq \, \in\, Y$ such that $L_{\wsq \, Y} = 0$.

\subsection{$q$-Whittaker Cauchy identities} 
From the Cauchy identities for $P_{\, Y} \ll \, \bx \, \rr$ and $Q_{\, Y} \ll \, \bx \, \rr$
\footnote{\,
Ch. VI, p. 324, Equation 4.13, and 
        p. 329, Equation 5.4, in \cite{macdonald.book}
},  
$W_{\, q \, Y} \ll \, \bx \, \rr$ and 
$W^{\, \prime}_{\, q \, Y} \ll\, \bx\, \rr$ satisfy the Cauchy identity, 
 
\begin{equation}
\sum_{\, Y} 
W_{\, q \, Y} \ll\, \bx\, \rr\, 
W^{\, \prime}_{\, q \, Y} \ll\, \by\, \rr  = 
\prod_{n = \, 0}^\infty
\ll 
\frac{1} 
     {1\, -\, \bx \, \by \, q^{\, n}}
\rr
\label{pqt.macdonald.cauchy.identity}
\end{equation}

\subsection{The structure constants}
From the product of two Macdonald functions
\footnote{\,
Ch. VI, p. 343, Equation $7.1^{\, \prime}$, in \cite{macdonald.book}
}, 
the product of two $q$-Whittaker functions can be expanded in the form, 

\begin{equation}
W_{\, q \,  Y_1} \ll\, \bx\, \rr \, 
W_{\, q \,  Y_2} \ll\, \bx\, \rr  = 
\sum_{\, Y_3} 
f_{\, Y_1 \, Y_2 \, Y_3} \, 
W_{\, q \,  Y_3} \ll\, \bx\, \rr, 
\label{product.01}
\end{equation}

which can be used as a definition of the $q$-dependent structure constants 
$f_{\, Y_1 \, Y_2 \, Y_3}$. Similarly, from the product of two dual Macdonald functions  
\footnote{\,
Ch. VI, p. 344, in \cite{macdonald.book} 
}, 
the product of two dual $q$-Whittaker functions can be expanded as, 

\begin{equation}
W^{\, \prime}_{\, q \,  Y_1} \ll\, \bx\, \rr \, 
W^{\, \prime}_{\, q \,  Y_2} \ll\, \bx\, \rr = 
\sum_{\, Y_3} 
f_{\, Y^{\, \prime}\, Y_2^{\, \prime} \, Y_3^{\, \prime}} \, 
W^{\, \prime}_{\, q \,  Y_3} \ll\, \bx\, \rr
\label{product.02}
\end{equation}

From Equations \ref{product.01} and \ref{product.02}, and the corresponding relations for 
Macdonald functions
\footnote{\,
Ch. VI, p. 344, Equation 7.3, in \cite{macdonald.book}
}, 

\begin{equation}
f_{\, Y_1^{\, \prime} \, Y_2^{\, \prime} \, Y_3^{\, \prime}} =
\ll
\frac{ 
b_{\, Y_3}
}{
b_{\, Y_1} \, 
b_{\, Y_2} 
} 
\rr\,
f_{\, Y_1 \, Y_2 \, Y_3} \, ,  
\end{equation}
	
where $b_{\, Y}$ is defined in Equation \ref{dual.macdonald}. Similarly
\footnote{\,
Ch. VI, p. 343, Equation 7.1, in \cite{macdonald.book}
}, 
the structure constant $f_{\, Y_1 \, Y_2 \, Y_3}$ can be written 
as an inner product, 

\begin{equation}
f_{\, Y_1 \, Y_2 \, Y_3} = 
\langle\,
W^{\, \prime}_{\, q \,  Y_3} \ll\, \bx\, \rr \, |\,
W_{\, q \,  Y_1} \ll\, \bx\, \rr \, 
W_{\, q \,  Y_2} \ll\, \bx\, \rr \,
\rangle
\end{equation}
	
\subsection{Skew $q$-Whittaker functions}
From the definitions of the skew Macdonald function 
$P_{Y_1 / Y_2} \ll \, \bx \, \rr$ and the skew dual Macdonald function
$Q_{Y_1 / Y_2} \ll \, \bx \, \rr$
\footnote{\,
Ch. VI, p. 344, Equation $7.6^{\, \prime}$, and 
Ch. VI, p. 344, Equation 7.5, 
in \cite{macdonald.book}
},
the skew $q$-Whittaker function 
$W_{\, q \, Y_1 / Y_2} \ll\, \bx\, \rr$ is defined in terms of the skew dual $q$-Whittaker 
function as, 
 
\begin{equation}
W_{\, q \, Y_1 / Y_2}       \ll\, \bx\, \rr  =
\ll \frac{ b_{\, Y_2}}
         { b_{\, Y_1}} 
\rr 
W^{\, \prime}_{\, q \, Y_1 / Y_2}       \ll\, \bx\, \rr, 
\end{equation}
	
while the skew dual $q$-Whittaker function $W^{\, \prime}_{\, q \, Y_1 / Y_2} \ll\, \bx\, \rr$ 
is defined in terms of the dual (non-skewed) $q$-Whittaker function as, 
	
\begin{equation}
W^{\, \prime}_{\, q \, Y_1 / Y_2}                 \ll\, \bx\, \rr  =
\sum_{\, Y_3} f_{\, Y_2\, Y_3 \, Y_1}  \, 
W^{\, \prime}_{\, q \,     Y_3}                 \ll\, \bx\, \rr 
\end{equation}

\subsection{Skew $q$-Whittaker Cauchy identities}
From the Cauchy identities for skew Macdonald functions
\footnote{\,
Ch. VI, p. 352, and p. 352, in \cite{macdonald.book}
}, 
the skew $q$-Whittaker functions satisfy the Cauchy identities, 
	
\begin{equation}
\sum_{\, Y} 
W^{         }_{\, q \,  Y    /   Y_1} \ll\,\bx\, \rr \, 
W^{\, \prime}_{\, q \, Y    /   Y_2} \ll\,\by\, \rr 
=
\prod_{     n = \, 0}^\infty
\ll 
\frac{1}
     {1\, -\, \bx \, \by \, q^{\, n}}
\rr
\sum_{\, Y}  
 \, 
W^{         }_{\, q \,  Y_2 / Y} \ll\,\bx\, \rr\, 
W^{\, \prime}_{\, q \,  Y_1 / Y} \ll\,\by\, \rr, 
\label{pqt.cauchy.identity.skew.01}
\end{equation}

and other Cauchy identities that involve skew  Hall-Littlewood functions that need 
not concern us here. 

\section{Twisted $q$-Whittaker functions}
\label{section.03}
\textit{We define an involution that we call \textit{\lq a twist\rq} that acts on the $q$-Whittaker 
functions to generate \lq twisted $q$-Whittaker functions\rq, then consider Cauchy identities that 
involve the $q$-Whittaker functions and their twisted versions. The reason why we need these specific 
Cauchy identities will be clear in the sequel.}

\subsection{A twist}

Consider the twist $\bi$, which acts on the power sum symmetric functions as, 

\begin{equation}
\bi \, . \, p_n \ll \bx \rr = \ll -1 \rr^{\, n-1} \, p_n \ll \bx \rr,
\quad
n = 1, 2, \cdots, 
\label{involution}
\end{equation}

which, \textit{in the absence of the Macdonald $q$ and $t$ parameters}, is identical to the involution 
$\omega$ in the theory of symmetric functions
\footnote{\,
Ch. I, p. 21 \cite{macdonald.book}.
},
and acts on Schur functions as,  

\begin{equation}
\bi    \, . \, s_{\lambda           } \ll \bx \rr =
\omega \, . \, s_{\lambda           } \ll \bx \rr =
                      s_{\lambda^{\,\prime}} \ll \bx \rr
\end{equation}

In the presence of Macdonald $q$ and $t$ parameters, the natural 
generalization of $\omega$ acts as,

\begin{equation}
\omega \, . \, p_n \ll \bx \rr = 
\ll -1 \rr^{\, n-1} \,  
\ll
\frac{
1 - q^{\, n}
}{
1 - t^{\, n}
}
\rr \, 
p_n \ll \bx \rr,
\quad
n \geq 1,
\label{involution.m}
\end{equation}

and acts on a $q$-Whittaker function with a parameter $q$ to give a Hall-Littlewood function
also with a parameter $q$ (rather than $t$) \cite{borodin.wheeler}. The action of our twist 
$\bi$ remains the same in the presence of a Macdonald parameter (or both), and acts on $q$-Whittaker 
functions to generate \textit{twisted $q$-Whittaker functions}. 
\subsection{Twisted $q$-Whittaker functions}
The involution $\bi$ acts on 
the      $q$-Whittaker functions $W^{         }_{\, q \,  Y}$ and 
the dual $q$-Whittaker functions $W^{\, \prime}_{\, q \, Y}$ 
to produce 
the twisted      $q$-Whittaker functions $W^{          \, \star}_{\, q \, Y}$ and 
the twisted dual $q$-Whittaker functions $W^{\, \prime \, \star}_{\, q \, Y}$ \, , 

\begin{equation}
\bi \, . \, W_{\, q \, Y} \ll \bx \rr = W^{\, \star}_{\, q \, Y} \ll \bx \rr, 
\quad 
\bi \, . \, W^{\, \prime}_{\, q \, Y} \ll \bx \rr = W^{\, \prime \, \star}_{\, q \, Y} \ll \bx \rr,
\label{twisted.whittaker}
\end{equation}

where $W^{\, \star}_{\, q \, Y} \ll \bx \rr$ and $W^{\, \prime \, \star}_{\, q \, Y} \ll \bx \rr$
are defined by expanding $W_{\, q \, Y} \ll \bx \rr$ and $W^{\, \prime}_{\, q \, Y} \ll \bx \rr$
in the power sum functions $p_{\, n} \ll \bx \rr$ and acting on the latter as in Equation 
\ref{involution}.

\subsubsection{Remark} 
In the limit $q \rightarrow 0$, both 
$W^{         }_{\, q \,  Y} \ll \bx \rr$ and 
$W^{\, \prime}_{\, q \, Y} \ll \bx \rr$ reduce to the same Schur function 
$s_{\, Y^{         }} \ll \bx \rr$ labelled by the Young diagram $Y$, and their twisted versions 
$W^{\,           \star}_{\, q \, Y} \ll \bx \rr$ and
$W^{\, \prime \, \star}_{\, q \, Y} \ll \bx \rr$ reduce to the same Schur function 
$s_{\, Y^{\, \prime}} \ll \bx \rr$ labelled by the transpose Young diagram $Y^{\, \prime}$. 

\subsection{More Cauchy identities} 
Starting from the identity, 

\begin{equation}
\exp 
\ll
\sum_{n=1}^{\infty} \,
\frac{1}{n} 
\frac{
1
}{
\ll 1 - q^{\, n} \rr
} \, 
p_n \ll \bx \rr 
p_n \ll \by \rr
\rr 
=
\prod_{\, n = 0}
\frac{
1
}{
\ll 1 - \, \bx \, \by \, q^{\, n} \rr
}, 
\end{equation}

and the identities obtained
by applying the involution \ref{involution} on the power sum functions $p_n \ll \bx \rr$ in 
the Cauchy identity \ref{pqt.cauchy.identity.skew.01},  

\begin{multline}
\exp 
\ll
\sum_{n=1}^{\infty} \,
\frac{1}{n} 
\frac{
1
}{
\ll 1 - q^{\, n} \rr
} \, 
\bi \, . \, p_n \ll \bx \rr 
            p_n \ll \by \rr
\rr 
=
\\ 
\exp 
\ll
\sum_{n=1}^{\infty} \,
\frac{1}{n} 
\frac{
1
}{
\ll 1 - q^{\, n} \rr
} \, 
            p_n \ll \bx \rr 
\bi \, . \, p_n \ll \by \rr
\rr
=
\prod_{n=0}^{\infty} \,
\ll 1 + \, \bx \, \by \, q^{\, n} \rr, 
\end{multline}

and,

\begin{equation}
\exp 
\ll
\sum_{n=1}^{\infty} \,
\frac{1}{n} 
\frac{
1
}{
\ll 1 - q^{\, n} \rr
} \, 
\bi \, . \, p_n \ll \bx \rr 
\bi \, . \, p_n \ll \by \rr
\rr 
=
\prod_{n=0}^{\infty} \,
\frac{
1
}{
\ll 1 - \, \bx \, \by \, q^{\, n} \rr
}, 
\end{equation}

we obtain, 

\begin{equation}
\sum_{\, Y} 
W^{\,  \star}_{\, q \, Y    /   Y_1} \ll\,\bx\, \rr \, 
W^{\, \prime}_{\, q \, Y    /   Y_2} \ll\,\by\, \rr
=
\prod_{     n = \, 0}^\infty
\ll 
1 + \, \bx \, \by \, q^{\, n}
\rr
\sum_{\, Y}  
 \, 
W^{\,  \star}_{\, q \, Y_2 / Y} \ll\,\bx\, \rr\, 
W^{\, \prime}_{\, q \, Y_1 / Y} \ll\,\by\, \rr \,,
\label{pqt.cauchy.identity.skew.02}
\end{equation}

\begin{equation}
\sum_{\, Y} 
W^{\,  \star}_{\, q \, Y    /   Y_1} \ll\,\bx\, \rr \, 
W^{\, \prime\, \star}_{\, q \, Y    /   Y_2} \ll\,\by\, \rr 
=
\prod_{     n = \, 0}^\infty
\ll \frac{1}{1 - \, \bx \, \by \, q^{\, n}} \rr
\sum_{\, Y}  
 \, 
W^{\,  \star}_{\, q \, Y_2 / Y} \ll\,\bx\, \rr\, 
W^{\, \prime \, \star}_{\, q \, Y_1 / Y} \ll\,\by\, \rr
\label{pqt.cauchy.identity.skew.03}
\end{equation}

Replacing $q$ with $q^{\prime}$, 

\begin{equation}
\sum_{\, Y} 
W^{         }_{\, q^\prime \,  Y    /   Y_1} \ll\,\bx\, \rr \, 
W^{\, \prime}_{\, q^\prime \, Y    /   Y_2} \ll\,\by\, \rr 
=
\prod_{     n = \, 0}^\infty
\ll 1\, -\, \bx \, \by \, q^{\, n+1}
\rr
\sum_{\, Y}  
 \, 
W^{         }_{\, q^\prime \,  Y_2 / Y} \ll\,\bx\, \rr\, 
W^{\, \prime}_{\, q^\prime \,  Y_1 / Y} \ll\,\by\, \rr, 
\label{pqt.cauchy.identity.skew.prime.01}
\end{equation}

\begin{equation}
\sum_{\, Y} 
W^{\,  \star}_{\, q^\prime \, Y    /   Y_1} \ll\,\bx\, \rr \, 
W^{\, \prime}_{\, q^\prime \, Y    /   Y_2} \ll\,\by\, \rr
=
\prod_{     n = \, 0}^\infty
\ll \frac{1}
{1 + \, \bx \, \by \, q^{\, n+1}}
\rr
\sum_{\, Y}  
 \, 
W^{\,  \star}_{\, q^\prime \, Y_2 / Y} \ll\,\bx\, \rr\, 
W^{\, \prime}_{\, q^\prime \, Y_1 / Y} \ll\,\by\, \rr \,,
\label{pqt.cauchy.identity.skew.prime.02}
\end{equation}

\begin{equation}
\sum_{\, Y} 
W^{\,  \star}_{\, q^\prime \, Y    /   Y_1} \ll\,\bx\, \rr \, 
W^{\, \prime\, \star}_{\, q^\prime \, Y    /   Y_2} \ll\,\by\, \rr 
=
\prod_{     n = \, 0}^\infty
\ll 1 - \, \bx \, \by \, q^{\, n+1} \rr
\sum_{\, Y}  
 \, 
W^{\,  \star}_{\, q^\prime \, Y_2 / Y} \ll\,\bx\, \rr\, 
W^{\, \prime \, \star}_{\, q^\prime \, Y_1 / Y} \ll\,\by\, \rr,
\label{pqt.cauchy.identity.skew.prime.03}
\end{equation}

where we used 

\begin{equation}
\exp
\ll 
\pm 
\sum_{n=1}^{\infty} \,
\frac{1}{1-q^{\, \prime \, n}} 
\frac{x^{\, n} y^{\, n}}{n} \rr = \exp
\ll 
\mp 
\sum_{n=1}^{\infty} \,
\frac{q^{\, n}}{1-q^{ \, n}} 
\frac{x^{\, n} y^{\, n}}{n} \rr =\prod_{n=0}^\infty 
\ll 1 - \, x \, y \, q^{\, n + 1}\rr^{\, \pm 1},
\end{equation}
to express the factors in Cauchy identities in terms of $q$. 

\section{$pqt$-Free bosons and $pqt$-vertex operators}
\label{section.04}
\textit{We recall basic facts related to Saito's $pqt$-Heisenberg algebras 
and $pqt$-vertex operators and note that Saito's deformation parameter $p$ 
appears on equal footing with Macdonald's parameter $q$. 
Setting $p=t$, all dependence on $p$ and on $t$ disappears, and the remaining 
parameter $q$ deforms Schur functions into $q$-Whittaker functions. 
}
\medskip 

\subsection{Two $pqt$-Heisenberg algebras}
Saito's free-boson realization of the elliptic extension of the Ding-Iohara-Miki algebra is based on 
\textit{two} $pqt$-Heisenberg algebras
\footnote{\,
Our notation is slightly different from, but equivalent to Saito's notation.
}, 
	
\begin{equation}
[a_m, a_n] = m 
\ll 1-p^{\, |m|} \rr
\ll 
\frac{1 - q^{\, |m|}}
     {1 - t^{\, |m|}}
\rr 
\delta_{m+n, 0} \, , 
\quad   
[b_m, b_n] = - m
\ll 
{1 - p^{\, \prime \, |m|}}
\rr 
\ll 
\frac{1 - q^{\, \prime \, |m|}}
     {1 - t^{\, \prime \, |m|}}
\rr 
\delta_{m+n, 0} \, , 
\label{two.heisenbergs}
\end{equation}

where $p^{\, \prime}, \cdots,$ stand for $1/p, \cdots$
$a_{\, n}$ and $b_{\, n}$, $n = 1, 2, \cdots$ act as 
creation     operators on the left  vacuum state $\langle \, 0 \, |$, and as 
annihilation operators on the right vacuum state $| \, 0 \, \rangle$, while
$a_{\, n}$ and $b_{\, n}$, $n = - 1, - 2, \cdots$ act as 
annihilation operators on the left  vacuum state $\langle \, 0 \, |$, and as 
creation     operators on the right vacuum state $| \, 0 \, \rangle$. 

\subsubsection{Remark} One should consider the writing of the second Heisenberg algebra in Equation \ref{two.heisenbergs} 
as short-hand notation. When performing computations, and particularly expansions in the various parameters, one should
work in terms of the variables $p < 1$, $q < 1$, and $t < 1$, rather than their inverses, which also makes it clear that the minus sign on the right hand side of the 
$b$-operator commutator is due to notation, and that both algebras have the same signature.

\subsection{Two-boson $pqt$-vertex operators} 
From the $pqt$-Heisenberg algebras, Saito defines two-boson $pqt$-vertex 
operators
\footnote{\,
Equations (3.9) and (3.8) respectively, in \cite{saito.01}, but in different notation. 
In particular, our 
$\phi_{\, +}^{\, p \, q \, t} \ll x \rr$ and 
$\phi_{\, -}^{\, p \, q \, t} \ll x \rr$ are 
Saito's 
$\phi^{\, \star} \ll p; x \rr$ and 
$\phi            \ll p; x \rr$, respectively, and 
our $b_n$ is Saito's $- b_n$.
}, 

\begin{multline} 
\phi_{\, \pm}^{\, p \, q \, t} \ll x \rr
=
\\
\exp
\ll 
\sum_{n = 1}^{\infty}
\ll
\frac{
1
}{
1 - p^{\, n}
}
\rr
\ll
\frac{
1 - t^{\, n}
}{
1 - q^{\, n}
}
\rr
\frac{x^{\, \mp \, n}}{n} \, 
a_{\, \pm \, n}
\rr 
\, 
\exp
\ll
\sum_{n = 1}^{\infty}
\ll
\frac{
1
}{
1 - p^{\, \prime \, n} 
}
\rr 
\ll
\frac{
1 - t^{\, \prime \, n}
}{
1 - q^{\, \prime \, n}
}
\rr 
\frac{
x^{\, \prime \, \mp \, n}
}{
n
} \, 
b_{\, \pm \, n}
\rr, 
\label{phi.plus.minus}
\end{multline}

where $x^{\, \prime}, \cdots$, stands for $1/x, \cdots$

\subsection{Two equivalent specializations} 
In Equations 
\ref{two.heisenbergs} and 
\ref{phi.plus.minus}, 
$p$ is Saito's elliptic deformation parameter, $q$ and $t$ are Macdonald parameters,  
$p$ and $q$ appear on equal footing, and the following two specializations lead to 
identical results, up to a renaming of the parameters. One can either 
\1 Set $q \rightarrow t$, so that all dependence on $q$ and on $t$ disappears, 
to work in terms of Schur functions that are deformed by Saito's elliptic deformation
parameter $p$, or 
\2 set $p \rightarrow t$, so that all dependence on $p$ and on $t$ disappears,  
to work in terms of Schur functions that are deformed by Macdonald's parameter $q$, 
which are $q$-Whittaker functions.  
In the following, we choose specialization \textbf{2} because that makes it clear that 
we are dealing with $q$-Whittaker functions whose properties follow directly from those 
of Macdonald functions. 

\section{$q$-Free bosons and $q$-vertex operators}
\label{section.05}
\textit{We set $q \rightarrow t$, in Saito's $pqt$-Heisenberg algebras and
$pqt$-vertex operators, so that all dependence on these parameters disappears, 
and we end up with pairs of expressions that depend on $q$ and on $1/q$.}

\subsection{Two $q$-Heisenberg algebras}
Consider the $q$-Heisenberg algebras, which are specializations of Saito's
$pqt$-Heisenberg algebras, 
	
\begin{equation}
[a_m, a_n] = m 
\ll 1-q^{\, |m|} \rr
\delta_{m+n, 0} \, , 
\quad 
[b_m, b_n] = - m
\ll 
{1 - q^{\, \prime \, |m|}}
\rr 
\delta_{m+n, 0} \, ,
\quad
[a_m, b_n] = 0, 
\label{pqt.heisenberg.algebras}
\end{equation}

and the $q$-vertex operators, which are specializations of Saito's $pqt$-vertex 
operators,

\begin{equation} 
\phi_{\, q \, \pm} \ll x \rr
=
\exp
\ll 
\sum_{n=1}^{\infty} \,
\ll
\frac{
1
}{
1 - q^{\, n}
}
\rr
\frac{x^{\, \mp \, n}}{n} \, 
a_{\, \pm \, n}
\rr
\, 
\exp
\ll
\sum_{n=1}^{\infty} \,
\ll
\frac{
1
}{
1 - q^{\, \prime \, n}
}
\rr 
\frac{x^{\, \prime \, \mp \, n}}{n} \, 
b_{\, \pm \, n}
\rr  
\label{phi}
\end{equation}

\subsubsection{Remark} 
Because the oscillators in Equation \ref{phi} have the same signs, when $\phi_{\, \pm}$ acts 
on a pair of Young diagrams $\ll Y_1, Y_2 \rr$, it generates a new pair of Young diagrams 
$\ll W_1, W_2 \rr$, such that 
$Y_1$ and $W_1$ are interlacing, and
$Y_2$ and $W_2$ are also be interlacing.

\subsection{One-boson $q$-vertex operators} 
We define the one-boson $q$-vertex operators
\footnote{\, 
The exponentials in definitions in Equation \ref{gamma.vertex.operators} differ by a minus sign 
from those typically used in the literature. The latter were followed in \cite{foda.wu.02}. 
This amounts to a redefinition of the creation and annihilation operators $a_n, n \in \ZZ$, 
which does not change the Heisenberg algebra.
}, 
		
\begin{equation}
\Gamma_{\, q \, a \, \pm} \ll x \rr =
\exp 
\ll 
\sum_{n = 1}^{\infty} 
\ll 
\frac{
1
}{
1 - q^{\, n}
}    
\rr 
\frac{
x^{\, \mp \, n}
}{
n} \, 
a_{\pm \, n} \rr,
\quad
\Gamma_{\, q^{\, \prime} \, b \, \pm} \ll x \rr =
\exp 
\ll \sum_{n = 1}^{\infty} 
\ll 
\frac{
1 
}{
1 - q^{\, \prime \, n}
} 
\rr 
\frac{
x^{\, \mp \, n}
}{
n
} \, b_{\pm \, n} \rr,	
\label{gamma.vertex.operators}
\end{equation}
		
which factorize the two-boson $q$-vertex operators as
\footnote{\, 
$\Gamma_{\, q               \, a \pm} \ll x \rr$ and
$\Gamma_{\, q^{\, \prime}   \, b \pm} \ll x \rr$ are defined in exactly the same way 
apart from the fact that 
$\Gamma_{\, q               \, a \pm} \ll x \rr$ depends on the $a$-oscillators and the parameter $q$, while
$\Gamma_{\, q^{\, \prime}   \, b \pm} \ll x \rr$ depends on the $b$-oscillators and the inverse parameter $1/q$.
},  
		
\begin{equation}
\phi_{\, + \, q} \ll x \rr =
\Gamma_{\, q       \, a \, +} \ll x        \rr
\Gamma_{\, q^{\, \prime} \, b \, +} \ll x^{\, \prime} \rr, 
\quad
\phi_{\, - \, q} \ll x \rr = 
\Gamma_{\, q       \, a \, -} \ll x        \rr
\Gamma_{\, q^{\, \prime} \, b \, - } \ll x^{\, \prime} \rr, 
\label{phi.vertex.operators}
\end{equation}
	
and satisfy the commutation relations
\footnote{\,
To prove the commutation relations in Equations \ref{gamma.commutator.01} and \ref{gamma.commutator.02}, 
we assume that $q < 1$, so that $q^{\, \prime} > 1$, so that all expansions and resummations must be 
made with respect to $q$. 
}, 
		
\begin{multline}
		\Gamma_{\, q \, a \, +} \ll x^{\, \prime} \rr 
		\Gamma_{\, q \, a \, -} \ll y \rr
=
\\
\exp
\ll 
\sum_{n = 1}^{\infty}
\frac{
1
}{
1-q^{\, n}
} 
\frac{
x^{\, n} 
y^{\, n}
}{
n
} \rr 
\Gamma_{\, q \, a \, -} \ll y                         \rr \, 
\Gamma_{\, q \, a \, +} \ll x^{\, \prime}             \rr
\, 
=
\frac{
1
}{
\ll x \,y \, | \, q \rr_{\infty}
} \, 
\Gamma_{\, q \, a \, -} \ll y \rr \, 
\Gamma_{\, q \, a \, +} \ll x^{\, \prime} \rr, 
\label{gamma.commutator.01}
\end{multline}

\begin{multline}
\Gamma_{\, q^{\, \prime} \, b \, +}\ll x \rr 
\Gamma_{\, q^{\, \prime} \, b \, -}\ll y^{\, \prime} \rr
= 
\\
\exp
\ll 
- \sum_{n = 1}^{\infty}
\frac{1}{1-q^{\, \prime \, n}} 
\frac{x^{\, \prime \, n} y^{\, \prime \, n}}{n} \rr 
\Gamma_{\, q^{\, \prime} \, b \, -} \ll y^{\, \prime} \rr \, 
\Gamma_{\, q^{\, \prime} \, b \, +} \ll x \rr
= 
\ll x^{\, \prime}   \, y^{\, \prime} \, | \, q^{\, \prime} \rr_{\infty} \, 
\Gamma_{\, q^{\, \prime} \, b \, -} \ll y^{\, \prime} \rr \, 
\Gamma_{\, q^{\, \prime} \, b \, +} \ll x             \rr
\, 
=
\\
\exp 
\ll 
\sum_{n = 1}^{\infty}
\frac{
q^{\, n}
}{
1 - q^{\, n}} 
\frac{
x^{\, \prime \, n} 
y^{\, \prime \, n}
}{
n
} 
\rr 
\Gamma_{\, q^{\, \prime} \, b \, -} \ll y^{\, \prime} \rr \, 
\Gamma_{\, q^{\, \prime} \, b \, +} \ll x             \rr
=		
\frac{1}{\ll x^{\, \prime}   \, y^{\, \prime}  \, q \, | \, q \rr_{\infty}} \, 
\Gamma_{\, q^{\, \prime} \, b \, -} \ll y^{\, \prime} \rr \, 
\Gamma_{\, q^{\, \prime} \, b \, +} \ll x             \rr, 
\label{gamma.commutator.02}
\end{multline}

so that, 

\begin{equation}
\langle \, 0 \, | \, 
\phi_{\, q \, +} \ll \bx^{\, \prime} \rr \, 
\phi_{\, q \, -} \ll \by             \rr \, 
| \, 0 \, \rangle  
=
\frac{
1
}{
\theta_{\, q} \ll \bx, \, \by \rr
}
\label{elliptic.macdonald.kernel}
\end{equation}

\subsection{An inverse vertex operator} 
We also need to define the inverse vertex operator,

\begin{equation}
\Gamma^{\, \prime}_{\, q^{\, \prime} \, b \, -} \ll x \rr =
\exp 
\ll 
\, - \, 
\sum_{n = 1}^{\infty} 
\ll 
\frac{
1 
}{
1 - q^{\, \prime \, n}
} 
\rr 
\frac{
x^{\, n}
}{
n
} \, b_{\, - \, n} \rr,	
\label{inverse.gamma.vertex.operators}
\end{equation}

which satisfies the commutation relation, 

\begin{multline}
\Gamma_{\, q^{\, \prime}\, b \, +}\ll x \rr 
\Gamma^{\, \prime}_{\, q^{\, \prime}\, b \, -}\ll y^{\, \prime} \rr
= 
\\
\exp
\ll 
+ \sum_{n = 1}^{\infty}
\frac{1}{1-q^{\, \prime \, n}} 
\frac{x^{\, \prime \, n} y^{\, \prime \, n}}{n} \rr 
\Gamma^{\, \prime}_{\, q^{\, \prime}\, b \, -}\ll y^{\, \prime} \rr \, 
\Gamma_{\, q^{\, \prime}\, b \, +}\ll x \rr 
= 
\frac{1}{\ll x^{\, \prime}   \, y^{\, \prime} \, | \, q^{\, \prime} \rr_{\infty}} \, 
\Gamma^{\, \prime}_{\, q^{\, \prime}\, b \, -}\ll y^{\, \prime} \rr \, 
\Gamma_{\, q^{\, \prime}\, b \, +}\ll x \rr 
\, 
=
\\
\exp 
\ll 
-\sum_{n = 1}^{\infty}
\frac{
q^{\, n}
}{
1 - q^{\, n}} 
\frac{
x^{\, \prime \, n} 
y^{\, \prime \, n}
}{
n
} 
\rr 
\Gamma^{\, \prime}_{\, q^{\, \prime}\, b \, -}\ll y^{\, \prime} \rr \, 
\Gamma_{\, q^{\, \prime}\, b \, +}\ll x \rr 
=		
\ll x^{\, \prime}   \, y^{\, \prime}  \, q \, | \, q \rr_{\infty} \, 
\Gamma^{\, \prime}_{\, q^{\, \prime}\, b \, -}\ll y^{\, \prime} \rr \, 
\Gamma_{\, q^{\, \prime}\, b \, +}\ll x \rr 
\label{gamma.commutator.02-minus}
\end{multline}

$\Gamma^{\, \prime}_{\, q^{\, \prime} \, b \, -} \ll x \rr$, 
as well as the vertex operator obtained from it by the action 
of the involution $\bi$, will 
be the only inverse vertex operators that we need.

\section{The power-sum/Heisenberg correspondence. The $a$-Heisenberg algebra}
\label{section.06}
\textit{Starting from the $q$-Whittaker Cauchy identities, we obtain identities 
that involve operator-valued $q$-Whittaker functions that act on $q$-Whittaker 
states. In this section, we consider the $a$-Heisenberg algebra that depends on 
$a_{\, \pm \, n}$ only.
}
\smallskip 
	
\subsection{An isomorphism}
	
Comparing the inner product of power-sum symmetric functions in the $q$-Whittaker basis, 
Equation \ref{young.diagram.power.sum.inner.product.a}, and the inner product of the right 
and left-states, 
Equation \ref{macdonald.orthogonality}, we deduce that the power-sum symmetric function 
basis is isomorphic to the Fock space spanned by the left-states $\langle \, a_{\, Y}\, |$, 
as well as that spanned by the right-states $|\, a_{\, Y}\, \rangle$, where $Y$ is a partition. 
In the left states, on which the operators $a_{ n}, n = 1, 2, \cdots,$ act as creation operators, 
we have the correspondence, 
	
\begin{equation}
p_n \ll \bx \rr \rightleftharpoons \, a_{ n}, \quad n \geq 1 
\label{power.sum.heisenberg.correspondence}
\end{equation}

In the right states, on which the operators $a_{ n}, n = -1, -2, \cdots,$ act as 
creation operators, we have the correspondence,

\begin{equation}
p_n \ll \bx \rr \rightleftharpoons \, a_{-n}, \quad n \geq 1  
\label{power.sum.heisenberg.correspondence.dual}
\end{equation}
	
\subsection{Operator-valued $q$-Whittaker functions and Cauchy identities}
Since the power-sum symmetric functions form a complete basis, we expand the $q$-Whittaker 
functions in terms of the power-sum symmetric functions, then formally replace the latter 
with Heisenberg generators to obtain operator-valued $q$-Whittaker functions that act either 
on left-or on right-states that are labeled by $q$-Whittaker functions..

\subsection{The action of vertex operators on $q$-Whittaker states}
From Equations \ref{an.exponential.is.a.product} and \ref{pqt.cauchy.identity.skew.01}, 

\begin{multline}
\exp 
\ll \sum_{n=1}^\infty 
\frac{1}{n} 
\ll \frac{1}{1\, -\,q^{\, n}} \rr 
p_n \ll\, \bx\, \rr\, 
p_n \ll\, \by\, \rr\,
\rr
\sum_{\, Y}
W_{\, q \, Y_1 / Y} \ll\, \bx     \, \rr\, 
W^{\, \prime}_{\, q \, Y_2 / Y} \ll\, \by     \, \rr
\\
= \sum_{\, Y}
W_{\, q \, Y / Y_2} \ll\, \bx     \, \rr\, 
W^{\, \prime}_{\, q \, Y / Y_1} \ll\, \by     \, \rr
\label{step.02}
\end{multline}

\subsubsection{The action of $\Gamma_{\, q \, a \, +}$ on a left-state}	
Using the power-sum/Heisenberg correspondence, Equation \ref{power.sum.heisenberg.correspondence.dual}, 
on $p_n \ll \bx \rr$ in Equation \ref{step.02}, we obtain the operator-valued $q$-Whittaker Cauchy 
identity,

\begin{multline}
\exp 
\ll 
\sum_{n=1}^\infty 
 \frac{1}{n} 
\ll \frac{1}{1\, -\,q^{\, n}} \rr
a_n\, 
p_n \ll\, \by\, \rr\, 
\rr
\sum_{\, Y}
W_{\, q \, Y_1 / Y} \ll\, \ba_{+}\, \rr\, 
W^{\, \prime}_{\, q \, Y_2 / Y} \ll\, \by    \, \rr
\\
= \sum_{\, Y}
W_{\, q \, Y / Y_2} \ll\, \ba_{+}\, \rr\, 
W^{\, \prime}_{\, q \, Y / Y_1} \ll\, \by    \, \rr 
\label{step.03}
\end{multline}

From the definition of the $\Gamma_{\, q \, a \, +}$ vertex operators, 
Equation \ref{gamma.vertex.operators}, 
	
\begin{equation}
\Gamma_{\, q \, a \, +} \ll \by^{\, \prime} \rr 
\sum_{\, Y}
W_{\, q \, Y_1 / Y} \ll\, \ba_{+}     \, \rr\, 
W^{\, \prime}_{\, q \, Y_2 / Y} \ll\, \by         \, \rr
= \sum_{\, Y}
W_{\, q \, Y / Y_2} \ll\, \ba_{+}     \, \rr\, 
W^{\, \prime}_{\, q \, Y / Y_1} \ll\, \by         \, \rr
\label{step.04}
\end{equation}

where we have used the notation in 
section \ref{products.on.sequences}. Acting with each side of Equation \ref{step.04.repeated} on a left vacuum state, 
	
\begin{equation}
\sum_{\, Y}
\langle\, 
W_{\, q \, Y_1 / Y}\, |\,
W^{\, \prime}_{\, q \, Y_2 / Y} \ll\, \by         \, \rr
\Gamma_{\, q \, a \, +} \ll \by^{\, \prime} \rr            
= \sum_{\, Y}
\langle\, 
W_{\, q \, Y / Y_2}\, |\, 
W^{\, \prime}_{\, q \, Y / Y_1} \ll\, \by\, \rr, 
\label{step.05}
\end{equation}

where $\langle\, W_{\, q \, Y_1 / Y_2}\,|$ is a state in the free-boson Fock space 
obtained by the action of the operator-valued $q$-Whittaker function labelled by 
the skew Young diagram $Y_1 / Y_2$, that is, by definition, 
	
\begin{equation}
\langle\, \emptyset\, |\, 
W_{\, q \, Y_1 / Y_2}  \ll\, \ba_{+}\, \rr\, 
= 
\langle\, W_{\, q \, Y_1 / Y_2}\, | 
\label{macdonald.action.01}
\end{equation}

Setting $Y_2 = \emptyset$ in Equation \ref{step.05}, we force $Y = \emptyset$ in the sum 
on the left hand side,
	
\begin{equation}
\langle\, 
W_{\, q \,  Y_1}\, |\,
\Gamma_{\, q \, a \, +} \ll \by^{\, \prime} \rr            
= \sum_{\, Y}
\langle\, 
W_{\, q \, Y}\, |\, 
W^{\, \prime}_{\, q \, Y / Y_1} \ll\, \by\, \rr
\label{step.06}
\end{equation}

\subsubsection{The action of\, $\Gamma_{\, q \, a \, -}$\, on a right-state}
Using the power-sum/Heisenberg correspondence, 
Equation \ref{power.sum.heisenberg.correspondence.dual}, on $p_n \ll \by \rr$ in Equation 
\ref{step.02}, we obtain the operator-valued $q$-Whittaker Cauchy identity,
	
\begin{multline}
\exp 
\ll \sum_{n=1}^\infty 
\frac{
1
}{
n
} 
\ll 
\frac{
1
}{
1\, -\,q^{\, n}
} 
\rr 
p_n \ll\, \bx\, \rr\, a_{-n}
\rr
\sum_{\, Y}
W_{\, q \, Y_1 / Y} \ll \, \bx     \, \rr\, 
W^{\, \prime}_{\, q \, Y_2 / Y} \ll \, \ba_{-}\, \rr
\\
= \sum_{\, Y}
W_{\, q \, Y / Y_2} \ll\, \bx     \, \rr\, 
W^{\, \prime}_{\, q \, Y / Y_1} \ll\, \ba_{-}\, \rr
\label{step.03.repeated}
\end{multline}
	
From the definition of the $\Gamma_{\, q \, a \, -}$ vertex operators in Equation \ref{gamma.vertex.operators}, 
	
\begin{equation}
\Gamma_{\, q \, a \, -} \ll \bx \rr 
\sum_{\, Y}
W_{\, q \, Y_1 / Y} \ll\, \bx     \, \rr\, 
W^{\, \prime}_{\, q \, Y_2 / Y} \ll\, \ba_{-}\, \rr
= \sum_{\, Y}
W_{\, q \, Y / Y_2} \ll\, \bx     \, \rr\, 
W^{\, \prime}_{\, q \, Y / Y_1} \ll\, \ba_{-}\, \rr
\label{step.04.repeated}
\end{equation}
	
Acting with each side of Equation \ref{step.04.repeated} on a right vacuum state, 
	
\begin{equation}
\Gamma_{\, q \, a \, -} \ll \bx \rr 
\sum_{\, Y}
W_{\, q \, Y_1 / Y} \ll\, \bx     \, \rr\, 
|\, 
W^{\, \prime}_{\, q \, Y_2 / Y} \, \rangle 
= \sum_{\, Y}
W_{\, q \, Y / Y_2} \ll\, \bx     \, \rr\, 
|\, 
W^{\, \prime}_{\, q \, Y / Y_1}\, \rangle, 
\label{step.05.repeated}
\end{equation}
		
where $|\, W^{\, \prime}_{\, q \, Y_1 / Y_2} \, \rangle$ is a state in 
the free boson Fock space obtained by the action of the operator-valued 
$q$-Whittaker function labelled by the skew Young diagram $Y_1 / Y_2$,
that is, by definition, 
	
\begin{equation}
W^{\, \prime}_{\, q \, Y_1 / Y_2}  \ll\, \ba_{-}\, \rr\, |\, \emptyset\, \rangle = 
|\, W^{\, \prime}_{\, q \, Y_1 / Y_2} \, \rangle 
\label{macdonald.action.02}
\end{equation}
	
Setting $Y_1 = \emptyset$ in Equation \ref{step.05.repeated}, we force $Y = \emptyset$ 
in the sum on the left hand side, 
	
\begin{equation}
\Gamma_{\, q \, a \, -} \ll \bx \rr 
|\, W^{\, \prime}_{\, q \,  Y_2} \, \rangle 
= \sum_{\, Y}
W_{\, q \, Y/Y_2} \ll\, \bx     \, \rr\, 
|\, 
W^{\, \prime}_{\, q \, Y}\, \rangle 
\label{step.06.repeated}
\end{equation}

\subsubsection{
The action of 
\, $\Gamma_{\, q \, a \, -}$ on a left-state, and
\, $\Gamma_{\, q \, a \, +}$ on a right-state}	
Using Equations 
\ref{step.06} and  
\ref{step.06.repeated}, then Equation 
\ref{pqt.cauchy.identity.skew.01},
	
\begin{multline}
\langle\, W_{\, q \,  Y_1}\, |\, 
\Gamma_{\, q \, a \, +} \ll\, \bx^{\, \prime} \rr
\Gamma_{\, q \, a \, -} \ll\, \by             \rr \, 
|\,       W^{\, \prime}_{\, q \, Y_2            }\, \rangle = 
\\
=\, \sum_{\, Y} 
W^{\, \prime}_{\, q \, Y / Y_1}\, \ll \bx \rr\, 
W_{\, q \, Y / Y_2            }\, \ll \by \rr 
= 
\prod_{n  \,      = \, 0}^\infty
\ll 
\frac{
1
}{
1\, -\, \bx \, \by \, q^{\, n}
}
\rr
\sum_{\, Y} 
W^{\, \prime}_{\, q \, Y_2 / Y} \ll \bx \rr\, 
W_{\, q \, Y_1 / Y} \ll \by \rr
\label{version.01}
\end{multline}
	
Using the $q$-vertex operator commutation relation, Equation
\ref{gamma.commutator.01}, then inserting a complete set of orthonormal 
states, 
	
\begin{multline} 
\langle\, 
W_{\, q \,  Y_1}  \, |\, 
\Gamma_{\, q \, a \, +} \ll \bx^{\, \prime} \rr 
\Gamma_{\, q \, a \, -} \ll \by            \rr \, 
|\, 
W^{\, \prime}_{\, q \,  Y_2}\, \rangle = 
\\ 
\prod_{n      \, = \, 0}^\infty
\ll 
\frac{
1
}{
1\, -\, \bx \, \by \, q^{\, n}
}
\rr
\langle\, 
W_{\, q \,  Y_1}  \, |\, 
\Gamma_{\, q \, a \, -} \ll \by             \rr \,  
\Gamma_{\, q \, a \, +} \ll \bx^{\, \prime} \rr 
\, | \, 
W^{\, \prime}_{\, q \,  Y_2}\, \rangle = 
\\ 
\prod_{n       \, = \, 0}^\infty
\ll 
\frac{
1
}{
1\, - \, \bx \, \by \, q^{\, n}}
\rr
\sum_{\, Y}
\langle \, W_{\, q \, Y_1} \, | \, 
\prod_{j = 1}^\infty  
\Gamma_{\, q \, a \, -} \ll y_j \rr 
\, | \,   W^{\, \prime}_{\, q \, Y}\, \rangle\, 
\langle\, W_{\, q \, Y}\, |\, 
\prod_{i\, = \, 1}^\infty  
\Gamma_{\, q \, a \, +} \ll x^{\, \prime}_i \rr \, | \, W^{\, \prime}_{\, q \, Y_2}\, \rangle
\label{version.02}
\end{multline}
	
Comparing the right hand sides of Equations \ref{version.01} and \ref{version.02},
we obtain the two identities, 

\begin{multline}
\ll 
\langle\, W_{\, q \,  Y_1}\, | 
\, 
\Gamma_{\, q \, a \, -} \ll \by \rr\,
\rr\, 
|\, W^{\, \prime}_{\, q \,  Y_2} \, \rangle 
=   
W_{\, q \, Y_1 / Y_2} \ll \by \rr, 
\\ 
\langle \, W_{\, q \,  Y_1} \, |\,
\ll 
\Gamma_{\, q \, a \, +} \ll \bx^{\, \prime} \rr\,
|\, W^{\, \prime}_{\, q \,  Y_2}\, \rangle\,
\rr 
=  
W^{\, \prime}_{\, q \, Y_2 / Y_1} \ll\, \bx \,\rr, 
\label{two.preliminary.identities}
\end{multline}

where the brackets indicate the state acted on by the vertex operators.
Since the states        $\langle\, W_{\, q \,  Y_1}\, |$ 
form a basis of left-states,  the states $|\, W^{\, \prime}_{\, q \,  Y_2}\, \rangle$
form a basis of right-states, and given the $q$-inner product 
Equation \ref{young.diagram.power.sum.inner.product.a}, 

\begin{equation}
\langle\, W_{\, q \,  Y_1}\, | 
\, 
\Gamma_{\, q \, a \, -} \ll \by \rr \, = 
\sum_{\, Y} 
\langle\, W_{\, q \, Y}\, |\, \alpha_{\, Y} \ll \by \rr, 
\quad 
\Gamma_{\, q \, a \, +} \ll \bx^{\, \prime} \rr\,
|\, W^{\, \prime}_{\, q \,  Y_2}\, \rangle\,
=
\sum_{\, Y}\, 
\beta_{\, Y} \ll \bx \rr\, 
|\, W^{\, \prime}_{\, q \,  Y_2}\, \rangle, 
\label{expanding}
\end{equation}

where $\alpha_{\, Y} \ll \by \rr$ and $\beta_{\, Y} \ll \bx \rr$ 
are expansion coefficients that carry the dependence on
$\bx$ and $\by$, while the expansion is in the set of Young 
diagrams $Y$. Using Equation \ref{two.preliminary.identities}, we determine 
$\alpha_{\, Y} \ll \by \rr$ and 
$ \beta_{\, Y} \ll \bx \rr$, 

\begin{multline}
\langle\, W_{\, q \,  Y_1}\, |\, 
\Gamma_{\, q \, a \, -} \ll \by \rr  =  
\sum_{\, Y} 
\langle\, W_{\, q \, Y}  \, |\, 
W_{\, q \, Y_1 / Y} \ll \by \rr, 
\\ 
\Gamma_{\, q \, a \, +}  \ll \bx^{\, \prime} \rr\, |\, W^{\, \prime}_{\, q \,  Y_1}\, \rangle =  
\sum_{\, Y}  
W^{\, \prime}_{\, q \, Y_1 / Y} \ll\, \bx \,\rr\, 
|\, W^{\, \prime}_{\, q \, Y} \rangle 
\label{two.identities}
\end{multline}

\subsection{The action of vertex operators on twisted $q$-Whittaker states}
Replacing the $q$-Whittaker functions with their twisted version, 
we obtain the following identities for the action of the vertex
operators, 

\begin{multline}
\ll 
\langle\, W^{\, \star}_{\, q  \,  Y_1}\, | 
\, 
\Gamma_{\, q \, a \, -} \ll \by \rr\,
\rr\, 
|\, W^{\, \prime\, \star}_{\, q  \,  Y_2} \, \rangle 
=   
W^{\, \star}_{\, q \, Y_1 / Y_2} \ll \by \rr, 
\\ 
\langle \, W^{\, \star}_{\, q  \,  Y_1} \, |\,
\ll 
\Gamma_{\, q \, a \, +}  \ll \bx^{\, \prime} \rr\,
|\, W^{\, \prime\, \star}_{\, q  \,  Y_2}\, \rangle\,
\rr 
=  
W^{\, \prime\, \star}_{\, q \, Y_2 / Y_1} \ll\, \bx \,\rr, 
\label{two.preliminary.identities.star}
\end{multline}

\begin{multline}
\langle\, W^{\, \star}_{\, q  \,  Y_1}\, |\, 
\Gamma_{\, q \, a \, -} \ll \by \rr  =  
\sum_{\, Y} 
\langle\, W^{\, \star}_{\, q \, Y}  \, |\, 
W^{\, \star}_{\, q \, Y_1 / Y} \ll \by \rr,
\\ 
\Gamma_{\, q \, a \, +}  \ll \bx^{\, \prime} \rr\, 
|\, W^{\, \prime\, \star}_{\, q  \,  Y_1}\, \rangle =  
\sum_{\, Y}  
W^{\, \prime\, \star}_{\, q \, Y_1 / Y} \ll\, \bx \,\rr\, 
|\, W^{\, \prime\, \star}_{\, q \, Y} \rangle
\label{two.identities.star}
\end{multline}

\section{The power-sum/Heisenberg correspondence. The $b$-Heisenberg algebra}
\label{section.07}
\textit{Starting from the $q$-Whittaker Cauchy identities, we obtain identities 
that involve operator-valued $q$-Whittaker functions that act on $q$-Whittaker 
states. In this section, we consider the $b$-Heisenberg algebra that depends on 
$b_{\, \pm \, n}$ only.
}
\smallskip 

\subsection{An isomorphism}
Due to the extra minus sign in the commutator of $b_{\pm n}$ bosons, the correspondence between the power-sum functions and oscillators is given by, 
	
\begin{equation}
p_n \ll \bx \rr \rightleftharpoons \, b_{ n}, \quad n \geq 1,  
\label{power.sum.heisenberg.correspondence.b}
\end{equation}

for the power-sum functions in the left states, and

\begin{equation}
p_n \ll \bx \rr \rightleftharpoons \, -b_{-n}, \quad n \geq 1,  
\label{power.sum.heisenberg.correspondence.dual.b}
\end{equation}

for the power-sum functions in the right states
\footnote{\,
The minus sign on the right hand side of Equation \ref{power.sum.heisenberg.correspondence.dual.b}
follows from the minus sign on the right hand side of the commutator of the $b$-oscillators, 
Equation \ref{two.heisenbergs}, and the fact that we write all expressions that involve $b$-oscillators in terms of $q^{\, \prime}$ to maintain similarity to those that involve $a$-oscillators.
}. 

\subsection{The action of operator-valued $q$-Whittaker functions on states}
From the Cauchy identity for skew $q$-Whittaker functions, Equation \ref{pqt.cauchy.identity.skew.01}, 
with $q$ replaced by $q^{\, \prime}$, and use Equation \ref{an.exponential.is.a.product} to obtain, 
	
\begin{multline}
\exp 
\ll \sum_{n=1}^\infty 
\frac{
1
}{
n
} 
\ll 
\frac{
1
}{
1 \, - \, q^{\, \prime\, n}
} 
\rr 
p_n \ll\, \bx\, \rr\, 
p_n \ll\, \by\, \rr\,
\rr
\sum_{\, Y}
W_{\,             q^{\, \prime} \, Y_1 / Y} \ll\, \bx     \, \rr\, 
W^{\, \prime}_{\, q^{\, \prime} \, Y_2 / Y} \ll\, \by     \, \rr
\\
= \sum_{\, Y}
W_{\,             q^{\, \prime} \, Y / Y_2} \ll\, \bx     \, \rr\, 
W^{\, \prime}_{\, q^{\, \prime} \, Y / Y_1} \ll\, \by     \, \rr
\label{step.02.b}
\end{multline}

\subsubsection{The action of $\Gamma_{\, q \, b \, +}$ on a left-state}	
Using the power-sum/Heisenberg correspondence, Equation \ref{power.sum.heisenberg.correspondence.dual.b}, 
on the power sum function $p_n \ll \bx \rr$, on both sides of Equation \ref{step.02.b}, we introduce free-boson mode 
operators that act as creation operators on a left-state, to obtain the operator-valued $q$-Whittaker Cauchy 
identity,

\begin{multline}
\exp 
\ll 
\sum_{n=1}^\infty 
\frac{
1
}{
n
} 
\ll 
\frac{
1
}{
1 \, - \, q^{\, \prime \, n}
} 
\rr
b_n \, 
p_n \ll \, \by \, \rr \, 
\rr
\sum_{\, Y}
W_{\,             q^{\, \prime} \, Y_1 / Y} \ll\, \bab_{+}\, \rr\, 
W^{\, \prime}_{\, q^{\, \prime} \, Y_2 / Y} \ll\, \by    \, \rr
\\
= \sum_{\, Y}
W_{\,             q^{\, \prime} \, Y / Y_2} \ll\, \bab_{+}\, \rr\, 
W^{\, \prime}_{\, q^{\, \prime} \, Y / Y_1} \ll\, \by    \, \rr, 
\label{step.03.b}
\end{multline}

From the definition of the $\Gamma_{\, q^{\, \prime} \, b \, +}$ vertex operators, 
Equation \ref{gamma.vertex.operators}, 
	
\begin{equation}
\Gamma_{\, q^{\, \prime} \, b \, +} \ll \by^{\, \prime} \rr 
\sum_{\, Y}
W_{\, q^{\, \prime} \, Y_1 / Y} \ll \, \bab_{+}     \, \rr\, 
W^{\, \prime}_{\, q^{\, \prime} \, Y_2 / Y} \ll\, \by         \, \rr
= \sum_{\, Y}
W_{\, q^{\, \prime} \, Y / Y_2} \ll\, \bab_{+}     \, \rr\, 
W^{\, \prime}_{\, q^{\, \prime} \, Y / Y_1} \ll\, \by         \, \rr
\label{step.04.b}
\end{equation}

Acting with each side of Equation \ref{step.04.b} on a left vacuum state, 
	
\begin{equation}
\sum_{\, Y}
\langle\, 
W_{\,             q^{\, \prime} \, Y_1 / Y}
\, | \,
W^{\, \prime}_{\, q^{\, \prime} \, Y_2 / Y} \ll\, \by         \, \rr
\Gamma_{\, q^{\, \prime} \, b \, +} \ll \by^{\, \prime} \rr            
= \sum_{\, Y}
\langle\, 
W_{\, q^{\, \prime} \, Y / Y_2}
\, | \, 
W^{\, \prime}_{\, q^{\, \prime} \, Y / Y_1} \ll\, \by\, \rr, 
\label{step.05.b}
\end{equation}

where $\langle\, W_{\, q^{\, \prime} \, Y_1 / Y_2}\,|$ is a state in 
the free-boson Fock space obtained by the action of the operator-valued 
$q$-Whittaker function labelled by the skew Young diagram $Y_1 / Y_2$,
	
\begin{equation}
\langle\, \emptyset\, |\, 
W_{\, q^{\, \prime} \, Y_1 / Y_2}  \ll\, \bab_{+}\, \rr\, 
= 
\langle\, W_{\, q^{\, \prime} \, Y_1 / Y_2}\, | 
\label{macdonald.action.01.b}
\end{equation}

Setting $Y_2 = \emptyset$ in Equation \ref{step.05.b}, we force $Y = \emptyset$ 
in the sum on the left hand side,
	
\begin{equation}
\langle\, 
W_{\, q^{\, \prime} \,  Y_1}\, |\,
\Gamma_{\, q^{\, \prime} \, b \, +} \ll \by^{\, \prime} \rr            
= \sum_{\, Y}
\langle\, 
W_{\, q^{\, \prime} \,  Y}\, |\, 
W^{\, \prime}_{\, q^{\, \prime} \, Y / Y_1} \ll \, \by \, \rr
\label{step.06.b}
\end{equation}

\subsubsection{The action of\, $\Gamma^{\, \prime}_{\, q \, b \, - }$\, on a right-state}
Using the power-sum/Heisenberg correspondence, 
Equation \ref{power.sum.heisenberg.correspondence.dual.b}, on $p_n \ll \by \rr$, to 
introduce free-boson mode operators that act as creation operators on a right-state, 
	
\begin{multline}
\exp 
\ll -\sum_{n=1}^\infty 
\frac{
1
}{
n
} 
\ll 
\frac{
1
}{
1 \, - \, q^{\, \prime\, n}
} 
\rr 
p_n \ll\, \bx\, \rr\,  b_{-n}
\rr
\sum_{\, Y}
W_{\, q^{\, \prime} \, Y_1 / Y} \ll\, \bx     \, \rr\, 
W^{\, \prime}_{\, q^{\, \prime} \, Y_2 / Y} \ll\,  \bab_{-}\, \rr
\\
= \sum_{\, Y}
W_{\, q^{\, \prime} \, Y / Y_2} \ll\, \bx     \, \rr\, 
W^{\, \prime}_{\, q \, Y / Y_1} \ll\,  \bab_{-}\, \rr
\label{step.03.repeated.b}
\end{multline}
	
From the definition of the $\Gamma^{\, \prime}_{\, q \, b \, - }$ vertex operator,  
Equation \ref{inverse.gamma.vertex.operators}, 
	
\begin{equation}
\Gamma^{\, \prime}_{\, q^{\, \prime} \, b \, - } \ll  \bx \rr
\sum_{\, Y}
W_{\, q^{\, \prime} \, Y_1 / Y} \ll\, \bx     \, \rr\, 
W^{\, \prime}_{\, q^{\, \prime} \, Y_2 / Y} \ll\,  \bab_{-}\, \rr
= \sum_{\, Y}
W_{\, q^{\, \prime} \, Y / Y_2} \ll\, \bx     \, \rr\, 
W^{\, \prime}_{\, q^{\, \prime} \, Y / Y_1} \ll\,  \bab_{-}\, \rr
\label{step.04.repeated.b}
\end{equation}
	
Acting with each side of Equation \ref{step.04.repeated.b} on a right vacuum state, 
	
\begin{equation}
\Gamma^{\, \prime}_{\, q^{\, \prime} \, b \, - } \ll  \bx \rr
\sum_{\, Y}
W_{\, q^{\, \prime} \, Y_1 / Y} \ll\, \bx     \, \rr\, 
|\, 
W^{\, \prime}_{\, q^{\, \prime} \, Y_2 / Y} \, \rangle 
= \sum_{\, Y}
W_{\, q^{\, \prime} \, Y / Y_2} \ll\, \bx     \, \rr\, 
|\, 
W^{\, \prime}_{\, q^{\, \prime} \, Y / Y_1}\, \rangle, 
\label{step.05.repeated.b}
\end{equation}
		
where $|\, W^{\, \prime}_{\, q^{\, \prime} \, Y_1 / Y_2} \, \rangle$ is a state in 
the free boson Fock space obtained by the action of the operator-valued $q$-Whittaker 
function labelled by the skew Young diagram $Y_1 / Y_2$,
	
\begin{equation}
W^{\, \prime}_{\, q^{\, \prime} \, Y_1 / Y_2}  \ll\,  \bab_{-}\, \rr\, |\, \emptyset\, \rangle = 
|\, W^{\, \prime}_{\, q^{\, \prime} \, Y_1 / Y_2} \, \rangle 
\label{macdonald.action.02.b}
\end{equation}
	
Setting $Y_1 = \emptyset$ in Equation \ref{step.05.repeated.b}, we force 
$Y = \emptyset$ in the sum on the left hand side, 
	
\begin{equation}
\Gamma^{\, \prime}_{\, q^{\, \prime} \, b \, - } \ll  \bx \rr 
|\, W^{\, \prime}_{\, q^{\, \prime} \,   Y_2} \, \rangle 
= \sum_{\, Y}
W_{\, q^{\, \prime} \, Y/Y_2} \ll\, \bx     \, \rr\, 
|\, 
W^{\, \prime}_{\, q^{\, \prime} \, \, Y}\, \rangle 
\label{step.06.repeated.b}
\end{equation}

\subsubsection{
The action of 
\, $\Gamma^{\, \prime}_{\, q \, b \, -}$ on a  left-state, and
\, $\Gamma_{\, q \, b \, +}$ on a right-state}	
Using Equations 
\ref{step.06.b} and  
\ref{step.06.repeated.b}, then Equation 
\ref{pqt.cauchy.identity.skew.01},
	
\begin{multline}
\langle\, W_{\, q^{\, \prime} \,   Y_1}\, |\, 
\Gamma_{\, q^{\, \prime} \, b \, +} \ll\,  \bx^{\, \prime} \rr
\Gamma^{\, \prime}_{\, q^{\, \prime} \, b \, -} \ll\,  \by             \rr\, 
|\,       W^{\, \prime}_{\, q^{\, \prime} \, Y_2            }\, \rangle = 
\\
\sum_{\, Y} 
W^{\, \prime}_{\, q^{\, \prime} \, Y / Y_1}\, \ll \bx \rr\, 
W_{\, q^{\, \prime} \, Y / Y_2            }\, \ll \by \rr 
= 
\prod_{n \, = \, 0}^\infty
\ll 
\frac{
1
}{
1\, -\, \bx \, \by \, q^{\, \prime \, n}
}
\rr
\sum_{\, Y} 
W^{\, \prime}_{\, q^{\, \prime} \, Y_2 / Y} \ll \bx \rr\, 
W^{         }_{\, q^{\, \prime} \, Y_1 / Y} \ll \by \rr
\label{version.01.b}
\end{multline}
	
Using the $q$-vertex operator commutation relation, Equation
\ref{gamma.commutator.02-minus}, the left hand side of Equation \ref{version.01.b} 
can be re-written as, 
	
\begin{multline} 
\langle\, 
W_{\, q^{\, \prime} \,   Y_1}  \, |\, 
\Gamma_{\, q^{\, \prime} \, b \, +} \ll  \bx^{\, \prime} \rr 
\Gamma^{\, \prime}_{\, q^{\, \prime} \, b \, -} \ll  \by             \rr \, 
|\, 
W^{\, \prime}_{\, q^{\, \prime} \,   Y_2}\, \rangle = 
\\ 
\prod_{n      \, = \, 0}^\infty
\ll 
\frac{
1
}{
1\, -\, \bx \, \by \, q^{\, \prime\, n}
}
\rr
\langle\, 
W_{\, q^{\, \prime} \,   Y_1}  
\, | \, 
\Gamma^{\, \prime}_{\, q^{\, \prime} \, b \, -} \ll  \by             \rr 
\Gamma_{\, q^{\, \prime} \, b \, +} \ll  \bx^{\, \prime} \rr \, 
\, | \, 
W^{\, \prime}_{\, q^{\, \prime} \,   Y_2}\, \rangle = 
\\ 
\prod_{n      \, = \, 0}^\infty
\ll 
\frac{
1
}{
1\, -\, \bx \, \by \, q^{\, \prime\, n}
}
\rr
\sum_{\, Y}
\langle\, W_{\, q^{\, \prime} \,   Y_1}
\, | \, 
\Gamma^{\, \prime}_{\, q^{\, \prime} \, b \, - } \ll  \by \rr
\, | \,  W^{\, \prime}_{\, q^{\, \prime} \,  Y}\, \rangle\, 
\langle\, W_{\, q^{\, \prime} \,  Y}
\, | \, 
\Gamma_{\, q^{\, \prime} \, b \, +} \ll \bx^{\, \prime} \rr \, |\, 
W^{\, \prime}_{\, q^{\, \prime} \,  Y_2}\, \rangle
\label{version.02.b}
\end{multline}
	
Comparing the right hand side of Equation \ref{version.01.b} and
that of Equation \ref{version.02.b},  

\begin{multline}
\ll 
\langle\, W_{\, q^{\, \prime} \,  Y_1} \, | 
\, 
\Gamma^{\, \prime}_{\, q^{\, \prime} \, b \, - } \ll \by \rr \,
\rr\, 
|\, W^{\, \prime}_{\, q^{\, \prime} \,  Y_2} \, \rangle 
=   
W_{\, q^{\, \prime} \, Y_1 / Y_2} \ll \by \rr, 
\\ 
\langle \, W_{\, q^{\, \prime} \,   Y_1} \, |\,
\ll 
\Gamma_{\, q \, b \, +}  \ll \bx^{\, \prime} \rr
\, | \, 
W^{\, \prime}_{\, q^{\, \prime} \,  Y_2}\, \rangle\,
\rr 
=  
W^{\, \prime}_{\, q^{\, \prime} \, Y_2 / Y_1} \ll\, \bx \,\rr, 
\label{two.preliminary.identities.b}
\end{multline}

where the brackets indicate the state acted on by the vertex operators.
Since the states        $\langle\, W_{\, q \,  Y_1}\, |$ 
form a basis of left-states,  and $|\, W^{\, \prime}_{\, q \,  Y_2}\, \rangle$
form a basis of right-states, and given the $q$-inner product 
Equation \ref{young.diagram.power.sum.inner.product.a} with $q$ replaced by $q^{\prime}$, 

\begin{equation}
\langle\, W_{\, q^{\, \prime} \,   Y_1}
\, | \, 
\Gamma^{\, \prime}_{\, q^{\, \prime} \, b \, - } \ll \by \rr \, = 
\sum_{\, Y} 
\langle\, W_{\, q^{\, \prime} \, \, Y} \, |\, \alpha_{\, Y} \ll \by \rr, 
\quad 
\Gamma_{\, q \, b \, +}  \ll \bx^{\, \prime} \,
|\, W^{\, \prime}_{\, q \,  Y_2} \, \rangle\,
\rr 
=
\sum_{\, Y}\, 
\beta_{\, Y} \ll \bx \rr\, 
|\, W^{\, \prime}_{\, q^{\, \prime} \,  Y_2}\, \rangle, 
\label{expanding.b}
\end{equation}

where $\alpha_{\, Y} \ll \by \rr$ and $\beta_{\, Y} \ll \bx \rr$ 
are expansion coefficients that carry the dependence on the variables
$\ll \bx \rr$ and $\ll \by \rr$, while the expansion is in the set of Young 
diagrams $Y$. Using Equation \ref{two.preliminary.identities}, we determine 
$\alpha_{\, Y} \ll \by \rr$ and 
$ \beta_{\, Y} \ll \bx \rr$, 

\begin{multline}
\langle\, W_{\, q^{\, \prime} \,   Y_1}\, |\, 
\Gamma^{\, \prime}_{\, q^{\, \prime} \, b \, - } \ll \by \rr  =  
\sum_{\, Y} 
\langle\, W_{q^{\, \prime} \, Y}  
\, | \, 
W_{\, q^{\, \prime} \, Y_1 / Y} \ll \by \rr, 
\\
\Gamma_{\, q^{\, \prime} \, b \, +}  \ll \bx^{\, \prime} \rr
\, | \, 
W^{\, \prime}_{\, q^{\, \prime} \,  Y_1}\, \rangle =  
\sum_{\, Y}  
W^{\, \prime}_{\, q^{\, \prime} \, Y_1 / Y} \ll\, \bx \,\rr
\, | \, 
W^{\, \prime}_{q^{\, \prime}\, Y} \rangle
\label{two.identities.b}
\end{multline}

\subsection{Twisted  $q$-Whittaker function identities}
Replacing the $q$-Whittaker functions with their twisted versions, 

\begin{multline}
\ll 
\langle\, W^{\, \star}_{\, q^{\, \prime} \,   Y_1}\, | 
\, 
\Gamma^{\, \prime}_{\, q^{\, \prime} \, b \, - } \ll \by \rr\,
\rr
\, | \, 
W^{\, \prime\, \star}_{\, q^{\, \prime} \,  Y_2} \, \rangle 
=   
W_{\, q^{\, \prime\, \star} \, Y_1 / Y_2} \ll \by \rr, 
\\ 
\langle \, W^{\, \star}_{\, q^{\, \prime} \,  Y_1} 
\, | \, 
\ll 
\Gamma_{\, q \, b \, +}  \ll \bx^{\, \prime} \rr
\, |\, 
W^{\, \prime\, \star}_{\, q^{\, \prime} \,  Y_2}\, \rangle \,
\rr 
=  
W^{\, \prime\, \star}_{\, q^{\, \prime} \, Y_2 / Y_1} \ll\, \bx \,\rr, 
\label{two.preliminary.identities.b.star}
\end{multline}

\begin{multline}
\langle\, W^{\, \star}_{\, q^{\, \prime} \,   Y_1}\, |\, 
\Gamma^{\, \prime}_{\, q^{\, \prime} \, b \, - } \ll \by \rr  =  
\sum_{\, Y} 
\langle\, W^{\, \star}_{q^{\, \prime} \,Y}  
\, | \, 
W^{\, \star}_{\, q^{\, \prime} \, Y_1 / Y} \ll \by \rr, 
\\
\Gamma_{\, q^{\, \prime} \, b \, +}  \ll \bx^{\, \prime} \rr
\, | \, 
W^{\, \prime\, \star}_{\, q^{\, \prime} \,  Y_1}\, \rangle =  
\sum_{\, Y}  
W^{\, \prime\, \star}_{\, q^{\, \prime} \, Y_1 / Y} \ll \, \bx \,\rr\, 
|\, W^{\, \prime\, \star}_{q^{\, \prime} \, Y} \rangle
\label{two.identities.b.star}
\end{multline}

Applying the twist $\bi$, Eqaution \ref{involution}, 
on $p_n \ll \by \rr$ on both sides of Equation \ref{two.preliminary.identities.b}, 
renaming the variables,  
and using,

\begin{equation}
\bi \, . \, \Gamma^{\, \prime}_{\, q^{\, \prime} \, b \, - } \ll   \, x \rr =
      \Gamma^{         }_{\, q^{\, \prime} \, b \, - } \ll - \, x \rr,
\end{equation}

\begin{multline}
\ll 
\langle\, W_{\, q^{\, \prime} \,  Y_1} \, | 
\, 
\Gamma_{\, q^{\, \prime} \, b \, - } \ll - \bx \rr \,
\rr\, 
|\, W^{\, \prime}_{\, q^{\, \prime} \,  Y_2} \, \rangle 
=   
W^{\star}_{\, q^{\, \prime} \, Y_1 / Y_2} \ll \bx \rr, 
\\
\langle\, W_{\, q^{\, \prime} \,   Y_1}\, |\, 
\Gamma_{\, q^{\, \prime} \, b \, - } \ll - \bx \rr  =  
\sum_{\, Y} 
\langle\, W_{q^{\, \prime} \, Y}  
\, | \, 
W^{\star}_{\, q^{\, \prime} \, Y_1 / Y} \ll \bx \rr 
\label{two.identities.b-twist}
\end{multline}

\section{$q$-Whittaker pairs and operator-valued identities}
\label{section.08}
\textit{We define pairs of $q$-Whittaker functions, and derive their Cauchy identities.
}
\medskip 

\subsection{$q$-Whittaker and twisted $q$-Whittaker pairs}

\begin{equation}
\bW_{\, \bfq \, \bY/\bR}   \ll \bx \rr = 
  W_{\, q             \, Y_1/R_1} \ll \bx \rr \, 
  W^{\, \star}_{\, q^{\, \prime} \, Y_2/R_2} \ll -\bx^{\, \prime} \rr, 
\quad
\bW^{\, \prime}_{\, \bfq \, \bY/\bR}   \ll \bx \rr = 
  W^{\, \prime}_{\, q             \, Y_1/R_1} \ll \bx \rr \, 
  W^{\, \prime}_{\, q^{\, \prime} \, Y_2/R_2} \ll \bx^{\, \prime} \rr
\end{equation}

\begin{equation}
\bW^{\, \star}_{\, \bfq \, \bY/\bR}   \ll \bx \rr = 
  W^{\, \star}_{\, q             \, Y_1/R_1} \ll \bx \rr \, 
  W_{\, q^{\, \prime} \, Y_2/R_2} \ll -\bx^{\, \prime} \rr,
\quad
\bW^{\, \prime\, \star}_{\, \bfq \, \bY/\bR}   \ll \bx \rr = 
  W^{\, \prime \, \star}_{\, q             \, Y_1/R_1} \ll \bx \rr \, 
  W^{\, \prime \, \star}_{\, q^{\, \prime} \, Y_2/R_2} \ll \bx^{\, \prime} \rr
\end{equation}

\subsection{Cauchy identities for $q$-Whittaker pairs.}

From the Cauchy identities in Equations 
\ref{pqt.cauchy.identity.skew.01}, 
\ref{pqt.cauchy.identity.skew.prime.02}, 
\ref{pqt.cauchy.identity.skew.03}, and applications of the involution $\bi$,  

\begin{equation}
\sum_{\bY}
\bW^{         }_{\, \bfq \, \bY/\bR} \ll \bx \rr \, 
\bW^{\, \prime}_{\, \bfq \, \bY/\bS} \ll \by \rr
=
\frac{
1
}{
\theta_q \ll \bx \, \by  \rr 
}
\sum_{\bW}
\bW^{         }_{\, \bfq \, \bS/\bW}   \ll \bx \rr 
\bW^{\, \prime}_{\, \bfq \, \bR/\bW}   \ll \by \rr
\label{double-normal-normal-Cauchy}
\end{equation}

\begin{equation}
\sum_{\bY}
\bW^{\,          \star}_{\, \bfq \, \bY/\bR} \ll \bx \rr \, 
\bW^{\, \prime\, \star}_{\, \bfq \, \bY/\bS} \ll \by \rr
=
\frac{
1
}{
\theta_q \ll  \bx \, \by  \rr 
}
\sum_{\bW}
\bW^{            \star}_{\, \bfq \, \bS/\bW}   \ll \bx \rr 
\bW^{\, \prime\, \star}_{\, \bfq \, \bR/\bW}   \ll \by \rr
\label{double-star-star-Cauchy}
\end{equation}

\begin{equation}
\sum_{\bY}
\bW^{                 }_{\, \bfq \, \bY/\bR} \ll \bx \rr \, 
\bW^{\, \prime\, \star}_{\, \bfq \, \bY/\bS} \ll \by \rr
=
\theta_q \ll - \bx \, \by  \rr 
\sum_{\bW}
\bW^{         }_{\, \bfq \, \bS/\bW}   \ll \bx \rr 
\bW^{\, \prime\, \star}_{\, \bfq \, \bR/\bW}   \ll \by \rr
\label{double-normal-star-Cauchy}
\end{equation}

\begin{equation}
\sum_{\bY}
\bW^{\star                 }_{\, \bfq \, \bY/\bR} \ll \bx \rr \, 
\bW^{\, \prime}_{\, \bfq \, \bY/\bS} \ll \by \rr
=
\theta_q \ll - \bx \, \by  \rr 
\sum_{\bW}
\bW^{  \star  }_{\, \bfq \, \bS/\bW}   \ll \bx \rr 
\bW^{\, \prime}_{\, \bfq \, \bR/\bW}   \ll \by \rr
\label{double-star-normal-Cauchy}
\end{equation}

\section{The elliptic vertex}
\label{section.09}
\textit{We construct an elliptic extension of the refined topological vertex using $q$-Whittaker functions, and a twisted version of the same vertex using 
twisted $q$-Whittaker functions.
}
\medskip 

We construct the elliptic vertex $\cE_{\, \bY_1 \bY_2 Y_3} \ll x, y \rr$ in five steps.

\subsection{Step \1 From $\bY_3$ to an infinite sequence of vertex operators} 
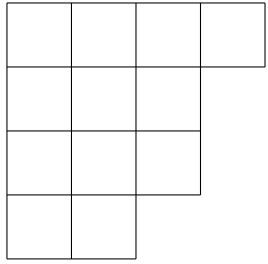
\begin{figure}
\begin{center}
\begin{tikzpicture}[scale=.8485]
\draw [thin] (-1.5,8.0)--(2.5,8.0);
\draw [thin] (-1.5,7.0)--(2.5,7.0);
\draw [thin] (-1.5,6.0)--(1.5,6.0);
\draw [thin] (-1.5,5.0)--(1.5,5.0);
\draw [thin] (-1.5,4.0)--(0.5,4.0);

\draw [thin] (-1.5,4.0)--(-1.5,8.0);
\draw [thin] (-0.5,4.0)--(-0.5,8.0);
\draw [thin] ( 0.5,4.0)--( 0.5,8.0);
\draw [thin] ( 1.5,5.0)--( 1.5,8.0);
\draw [thin] ( 2.5,7.0)--( 2.5,8.0);

\end{tikzpicture}
\end{center}
\caption{ \it A Young diagram $Y = \ll 4, 3, 3, 2 \rr$ 
}
\label{young.diagram}
\end{figure}

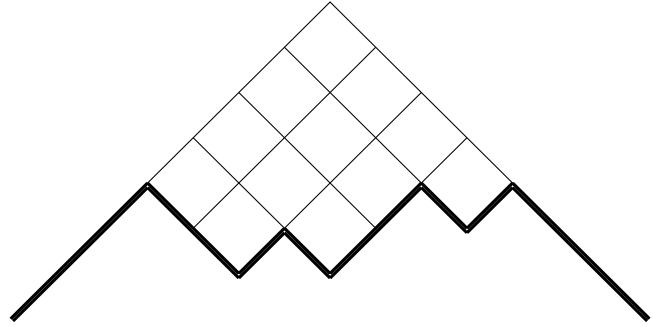
\begin{figure}
\begin{center}
\begin{tikzpicture}[scale=.6]
\draw [thin] (-1.5,1.0)--(5.5,8.0)--(12.5,1.0);
\draw [thin] ( 2.5,3.0)--(6.5,7.0);
\draw [thin] ( 3.5,2.0)--(7.5,6.0);
\draw [thin] ( 5.5,2.0)--(8.5,5.0);
\draw [thin] ( 8.5,3.0)--(9.5,4.0);
\draw [thin] ( 1.5,4.0)--(3.5,2.0);
\draw [thin] ( 2.5,5.0)--(5.5,2.0);
\draw [thin] ( 3.5,6.0)--(6.5,3.0);
\draw [thin] ( 4.5,7.0)--(8.5,3.0);

\draw [very thick] (-1.5,1.0)--(1.5,4.0);
\draw [very thick] (-1.45,0.95)--(1.5,3.9);

\draw [very thick] ( 1.5,4.0)--(3.5,2.0);
\draw [very thick] ( 1.5,3.9)--(3.5,1.9);

\draw [very thick] ( 3.5,2.0)--(4.5,3.0);
\draw [very thick] ( 3.5,1.9)--(4.5,2.9);

\draw [very thick] ( 4.5,3.0)--(5.5,2.0);
\draw [very thick] ( 4.5,2.9)--(5.5,1.9);

\draw [very thick] ( 5.5,2.0)--(7.5,4.0);
\draw [very thick] ( 5.5,1.9)--(7.5,3.9);

\draw [very thick] ( 7.5,4.0)--(8.5,3.0);
\draw [very thick] ( 7.5,3.9)--(8.5,2.9);

\draw [very thick] ( 8.5,3.0)--(9.5,4.0);
\draw [very thick] ( 8.5,2.9)--(9.5,3.9);

\draw [very thick] ( 9.5,4.0)--(12.5,1.0);
\draw [very thick] ( 9.5,3.9)--(12.45,0.95);

\end{tikzpicture}
\end{center}
\caption{ \it A tilted Young diagram $Y$ and its infinite profile indicated by 
a heavy line.
}
\label{tilted.young.diagram}
\end{figure}

\begin{figure}
\begin{center}
\begin{tikzpicture}[scale=.6]
\draw [thin] (-1.5,1.0)--(5.5,8.0)--(12.5,1.0);
\draw [thin] ( 2.5,3.0)--(6.5,7.0);
\draw [thin] ( 3.5,2.0)--(7.5,6.0);
\draw [thin] ( 5.5,2.0)--(8.5,5.0);
\draw [thin] ( 8.5,3.0)--(9.5,4.0);
\draw [thin] ( 1.5,4.0)--(3.5,2.0);
\draw [thin] ( 2.5,5.0)--(5.5,2.0);
\draw [thin] ( 3.5,6.0)--(6.5,3.0);
\draw [thin] ( 4.5,7.0)--(8.5,3.0);

\draw [very thick] (-1.5,1.0)--(1.5,4.0);
\draw [very thick] (-1.45,0.95)--(1.5,3.9);

\draw [very thick] ( 1.5,4.0)--(3.5,2.0);
\draw [very thick] ( 1.5,3.9)--(3.5,1.9);

\draw [very thick] ( 3.5,2.0)--(4.5,3.0);
\draw [very thick] ( 3.5,1.9)--(4.5,2.9);

\draw [very thick] ( 4.5,3.0)--(5.5,2.0);
\draw [very thick] ( 4.5,2.9)--(5.5,1.9);

\draw [very thick] ( 5.5,2.0)--(7.5,4.0);
\draw [very thick] ( 5.5,1.9)--(7.5,3.9);

\draw [very thick] ( 7.5,4.0)--(8.5,3.0);
\draw [very thick] ( 7.5,3.9)--(8.5,2.9);

\draw [very thick] ( 8.5,3.0)--(9.5,4.0);
\draw [very thick] ( 8.5,2.9)--(9.5,3.9);

\draw [very thick] ( 9.5,4.0)--(12.5,1.0);
\draw [very thick] ( 9.5,3.9)--(12.45,0.95);

\foreach \i in {0,...,6} 
{
\draw [dashed, red] (\i - .5, \i + 1.5)--(\i - .5, 0);
}
\foreach \j in {7,...,12} 
{
\draw [dashed, red](\j -.5, 13.5 -\j)--(\j - .5, 0);
}
\node [left]  at (-1.3,0) {$\cdots$};
\node [right] at (12.3,0) {$\cdots$};
\foreach \x in {0,...,2}
{
\draw                 (\x- 1,0) circle (0.3);
\draw [fill=black!50] (\x+10,0) circle (0.3);
}

\draw [fill=black!50] (2,0) circle (0.3);
\draw [fill=black!50] (3,0) circle (0.3);
\draw (4,0) circle (0.3);
\draw [fill=black!50] (5,0) circle (0.3);
\draw (6,0) circle (0.3);
\draw  (7,0) circle (0.3);
\draw [fill=black!50] (8,0) circle (0.3);
\draw (9,0) circle (0.3);

\foreach \a in {6, 5, ..., 1} 
{
\node [below] at (- \a + 5.8, -.3) {$- \a$};
}
\foreach \a in {0, 1, ..., 5} 
{
\node [below] at (  \a + 6.0, -.3) {$  \a$};
}				
\end{tikzpicture}
\end{center}
\caption{ \it The tilted Young diagram, its infinite profile, and the corresponding 
Maya diagram, which gives the Young diagram/Maya diagram correspondence for 
$Y = \ll 4, 4, 3, 1 \rr$. The integer below a stone is its position in the Maya diagram.
The apex of the inverted Young diagram is located between positions $-1$ and $0$.
}
\label{young.maya}
\end{figure}
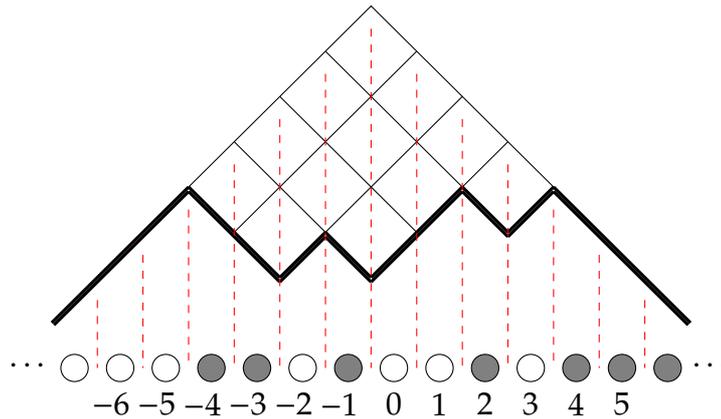

We consider the \textit{finite} Young diagram $Y_3$ that labels the preferred leg of the 
vertex that we wish to construct, see Figure \ref{young.diagram}, position it as in Figure 
\ref{tilted.young.diagram}, and consider its \textit{infinite profile}, which consists of 
upward and downward segments $\ll \diagup, \diagdown \rr$, that (scanning the profile from 
$- \infty$ all the way to the left, to $\infty$ all the way to the right) are all upward 
sufficiently far to the left, and all downward sufficiently far to the right, as indicated 
in Figure \ref{young.maya}. We map this infinite profile to a Maya diagram 
\cite{miwa.jimbo.date.book}, that consists white and black stones $\ll \Circle, \CIRCLE \rr$, 
then we map the latter to an infinite sequence of $q$-vertex operators, that we denote by 
$\prod_{Maya \ll Y_3 \rr} \phi_{\, q \, \pm}$
\footnote{\,
The vertex operators $\phi_{\, q \, \pm}$ are defined in Equation \ref{phi.vertex.operators} in terms of one-boson vertex operators
that are \textit{not inverted}. 
}, 
using the bijections
\footnote{\,
We could have skipped the intermediate step of mapping to a Maya diagram, but we prefer to 
keep because it can be useful in related contexts.}, 

%FIGURE.01  

%FIGURE.02

%FIGURE.03  

\begin{equation}
\diagup   \rightleftharpoons \Circle \rightleftharpoons \phi_{\, q \, -}, 
\quad
\quad
\diagdown \rightleftharpoons \CIRCLE \rightleftharpoons \phi_{\, q \, +}  
\end{equation}

\subsection{Step \2 Choosing the arguments of the vertex operators}
We choose the arguments of the $q$-vertex operators to be,

\begin{equation}
\phi_{\, q \, +} \ll x^{  - i}\, y^{  y^{\, \prime}_{3, i}            }   \rr \, , \quad 
\phi_{\, q \, -} \ll y^{j - 1}\, x^{- y_{3, j}}   \rr, 
\label{full.arguments}
\end{equation}

where 
$y_{3, i}$ is the length of the $i$-column of the Young diagram $Y_3$ that labels the 
preferred leg of the vertex, and 
$y^{\, \prime}_{3, j}$ is the length of the $j$-column of the transpose Young diagram 
$Y^{\, \prime}_3$.

\subsubsection{Example} The Young diagram/Maya diagram correspondence in 
Figure \ref{young.maya} leads to the vertex-operator sequence, 

\begin{multline}
\ll 
\prod_{Maya \ll Y_3 \rr} \phi_{\, q \, \pm}
\rr 
=
\cdots 
\phi_{\, q \, +}\ll x^{-5}     \rr
\phi_{\, q \, -}\ll x^{-4}     \rr
\phi_{\, q \, -}\ll y     \, x^{-4}    \rr
\phi_{\, q \, +}\ll x^{-4}\, y^{ 2}    \rr
\phi_{\, q \, -}\ll y^2   \, x^{-3}    \rr
\\
\phi_{\, q \, +}\ll x^{-3}\, y^3\rr
\phi_{\, q \, +}\ll x^{-2}\, y^3\rr
\phi_{\, q \, -}\ll y^3   \, x^{-1} \rr
\phi_{\, q \, +}\ll x^{-1}\, y^4 \rr
\phi_{\, q \, -}\ll          y^4 \rr
\cdots
\label{vertex.operator.sequence}
\end{multline}

\subsection{Step \3 From the infinite sequence of vertex operators to an expectation value} 
We evaluate the sequence 
$\prod_{Maya \ll Y_3 \rr} \phi_{\, q \, \pm}$ 
between a left-state that corresponds to a $q$-Whittaker pair labelled by a pair of Young diagrams $\bY_1$
\footnote{\,
The $q$-Whittaker pair that labels the left states were created by the action of pairs of one-boson vertex operators, one of which, that which depends on $b$-oscillators is inverted.

}

and a right-state 
labelled by a dual $q$-Whittaker pair labelled by a pair of Young diagrams $\bY_2$,

\begin{equation}
\cE^{\, unnorm}_{\, \bY_1 \, \bY_2 \, Y_3} \ll x, y \rr 
=
\langle\, 
\bW_{\, \bfq \, \bY_1} 
\, | \,
\ll 
\prod_{ \textit{Maya} \ll Y_3 \rr}
\phi_{\, q \, \pm} 
\rr 
\, |\, 
\bW^{\, \prime}_{\, \bfq \, \bY_2} \rangle
\label{unnormalized.elliptic.vertex}
\end{equation} 

To evaluate the expectation value in Equation \ref{unnormalized.elliptic.vertex}, we 
\lq order\rq\, the sequence in Equation \ref{unnormalized.elliptic.vertex}
\textit{via} an infinite number of commutations that put 
all $\phi_{\, q \, +}$ vertex operators on the right, and 
all $\phi_{\, q \, -}$ vertex operators on the left. 
From Equation \ref{elliptic.macdonald.kernel}, 

\begin{multline}
\phi_{\, q \, +} \ll x^{  - i}\, y^{  y^{\, \prime}_{3, i}} \rr  \,
\phi_{\, q \, -} \ll y^{j - 1}\, x^{- y_{3, j}} \rr =
\\
\prod_{m, \, n \, = \, 1}^\infty 
\ll 
\frac{
1
}{
\theta_{\, q} 
\ll
x^{\, - \, y_{\, 3, \, n} + m} \, y^{- y^{\, \prime \, +}_{\, 3, \, m} + n}
\rr
} 
\rr
\phi_{\, q \, -} \ll  y^{\, j \, - \, 1} \, x^{\, - \, y^{         }_{\, 3, \, j}} \rr\,
\phi_{\, q \, +} \ll  x^{\,      - \, i} \, y^{\,      y^{\, \prime}_{\, 3, \, i}} 
\rr, 
\label{q.t.vertex.operator.commutation.relation.with.arguments}
\end{multline}

where $y^+_{\, 3,\, i} = y_{\, 3,\, i} + 1$, and $y_{\, 3, \, i}$ is the length of 
the $i$-column in $Y_3$. Since  
$\phi_{\, q \, +}$ is attached to a segment 
$\diagup$ in the extended profile of $Y_3$, and 
$\phi_{\, q \, -}$ is attached to an adjacent segment 
$\diagdown$ to the right of the former, 
the commutation relation, 
Equation \ref{q.t.vertex.operator.commutation.relation.with.arguments}
describes replacing the adjacent pair 
$\diagup   \diagdown$ with the pair 
$\diagdown   \diagup$, adding a cell to $Y_3$, to generate a Young diagram that is 
larger by one cell. The exponents that appear in the factor on the right hand side 
of Equation \ref{q.t.vertex.operator.commutation.relation.with.arguments} have simple 
interpretations,

\begin{equation}
y_{\, 3, \, i}             - j = L_{\, \wsq}, 
\quad 
y_{\, 3, \, j}^{\, \prime} - i = A_{\, \wsq}, 
\end{equation}

where $A_{\wsq}$ and $L_{\wsq}$ are the arm-length and the leg-length of the cell 
$\wsq$ that is added to $Y_3$ \textit{ via} the commutation in Equation
\ref{q.t.vertex.operator.commutation.relation.with.arguments}, to generate a larger 
Young diagram, that is $\wsq \notin Y_3$. Inserting the sequence 
$\prod_{Maya \ll Y_3 \rr} \phi_{\, q \, \pm}$ 
between a left-state 
$\langle\, \bW_{\, \bfq \, \bY_1}\, |$, and a right-state 
$|\, \bW^{\, \prime}_{\, \bfq \, \bY_2}\, \rangle$, then commuting the (infinitely-many) 
$\phi_{\, q \, +}$ vertex operators to the right of the  
$\phi_{\, q \, -}$ vertex operators, 

\begin{multline}
\langle\, 
\bW_{\, \bfq \, \bY_1}\, |\, 
\ll \prod_{i=1}^\infty \phi_{\, q \, +} \ll x^{  - i}\, y^{  y^{\, \prime}_{            3, i}} \rr  \rr \,
\ll \prod_{j=1}^\infty \phi_{\, q \, -} \ll y^{j - 1}\, x^{- y_{3, j}} \rr  \rr \, 
| \, 
\bW^{\, \prime}_{\, \bfq \, \bY_2}\, 
\rangle\, 
=
\\
\ll 
\prod_{\wsq \notin Y_3} 
\frac{
1
}{
\theta_{\, q} \, 
\ll x^{- L_\wsq}\, y^{- A^{\, +}_\wsq} \, q^{\, n} \rr 
} 
\rr \, 
\langle\, \bW_{\, \bfq \, \bY_1}\, |
\ll \prod_{j=1}^\infty \phi_{\, q \, -} \ll  y^{j - 1}\, x^{- y_{3, j}} \rr \rr\,
\ll \prod_{i=1}^\infty \phi_{\, q \, +} \ll  x^{  - i}\, y^{  y^{\, \prime}_{            3, i}} \rr \rr
|\, \bW^{\, \prime}_{\, \bfq \, \bY_2}\, \rangle
\label{result.of.step.03}
\end{multline} 

\subsection{Step \4 From an expectation value to the unnormalized elliptic vertex}
Using the identities, 

\begin{equation}
\langle \, \bW_{\, \bfq \,  \bY_1} \, | \, 
\phi_{\, q \, -} \ll \by \rr  =  
\sum_{\, \bY}  
\langle \, \bW_{\, \bfq \, \bY}  \, | \, 
\bW_{\, \bfq \, \bY_1 / \bY} \ll \by \rr, 
\, \, 
\phi_{\, q \, +}  \ll \bx^{\, \prime} \rr \, | \, \bW^{\, \prime}_{\, \bfq \, \bY_1} \, \rangle =  
\sum_{\, \bY}  
\bW^{\, \prime}_{\, \bfq \, \bY_1 / \bY} \ll \, \bx \, \rr 
\, | \, \bW^{\, \prime}_{\, \bfq \, \bY} \rangle, 
\label{two.identities.combine}
\end{equation}

in Equation \ref{result.of.step.03}, we obtain the \textit{unnormalized} elliptic vertex, 

\begin{multline}
\cE^{\, unnorm}_{\, \bY_1 \, \bY_2 \, Y_3} \ll \bx, \, \by \rr = 
\langle \, \bW_{\, \bfq \, \bY_1} \, | \, 
\ll \prod_{i=1}^\infty \phi_{\, q +} \ll x^{  - i}\, y^{  y^{\, \prime}_{            3, i}} \rr \rr\,
\ll \prod_{j=1}^\infty \phi_{\, q \, -} \ll y^{j - 1}\, x^{- y_{3, j}} \rr \rr\, 
|\, \bW^{\, \prime}_{\, \bfq \, \bY_2}\, \rangle\, 
=
\\
\ll 
\prod_{\wsq \notin Y_3} 
\frac{
1
}{
\theta_{\, q} 
\ll x^{- L_\wsq} \, y^{- A^{\, +}_\wsq} \rr 
} 
\rr 
\,
\sum_{\, \bY}
\bW_{\, \bY_1 / \bY} \ll y^{\, \bj - 1}\, x^{\, - Y_3} \rr 
\bW^{\, \prime}_{\, \bY_2 / \bY} \ll x^{\, \bi    }\, y^{\, - Y_3^{\, \prime} } \rr, 
\label{macdonald.top.vertex.step.02}
\end{multline} 

where 
$\bi = \ll 1, 2, \cdots \rr$, 
$\bj = \ll 1, 2, \cdots \rr$, and the arguments in 
$\bW_{\, \bY_1 / Y} \ll y^{\, \bj    - 1}\, x^{\, -Y_3^{\, \prime}} \rr$ and 
$\bW^{\, \prime}_{\, \bfq \, \bY_2 / Y} \ll x^{\, \bi       }\, y^{\, -Y_3            } \rr$, 
should be understood in the sense of Section \ref{sequences}.

\subsection{Step \5 From the unnormalized to the normalized elliptic topological 
vertex}
To normalize the expression in Equation \ref{macdonald.top.vertex.step.02} such 
that $\cE_{\, \pmb{\emptyset} \, \pmb{\emptyset} \, \emptyset} = 1$, we divide 
it by the elliptic version of the $xy$-refined $q$-MacMahon partition function,
that is, 

\begin{equation}
\cM \ll x, y, q \rr 
= 
\prod_{i,\,  j \, = \, 1}^\infty \, 
\frac{
1
}{
\theta_{\, q} 
\ll 
x^{ i}\, y^{j-1}
\rr 
}, 
\label{x.y.q.t.macmahon.function.01}
\end{equation} 

Using the identity, 

\begin{equation}
\ll 
\prod_{\wsq \notin Y_3} 
\frac{
1
}{
1\, -\, x^{- L_\wsq}\, y^{- A^{\, +}_\wsq}\,q^{\, n}
} 
\rr 
\, 
\ll
\prod_{i, j = 1}^\infty 
\frac{
1
}{
1 - x^i \, y^{\,j-1}\,q^{\, n}
}
\rr^{-1}
=
\ll 
\prod_{\wsq \in Y_3}
\frac{
1
}{
1 -\,x^{\, L^+_\wsq} \, y^{\,A_\wsq} \, q^{\, n}
}
\rr,  
\end{equation}

which follows from Equations \textbf{2.8} and \textbf{2.11} in \cite{awata.kanno.02}. 
The final, normalised elliptic vertex $\cE_{\, \bY_1\,\bY_2\, Y_3} \ll x, y \rr$ is,

\begin{equation}
\cE_{\, \bY_1\,\bY_2\, Y_3} \ll x, y \rr 
=
\frac{
\cE^{\, unnorm}_{\bY_1     \,\bY_2     \,Y_3} \ll x, y \rr
}{
\cE^{\, unnorm}_{\emptyset\,\emptyset\,\emptyset} \ll x, y \rr
}, 
\label{normalized.macdonald.vertex}
\end{equation}

where, 

\begin{empheq}[box=\fbox]{equation}
\cE_{\, \bY_1 \, \bY_2 \, Y_3} \ll x, y \rr 
= 
\prod_{\wsq \in Y_3}
\frac{
1
}{
\theta_q \ll x^{\,L^+_{\wsq \, Y_3}} \, y^{\,A_{\wsq \, Y_3}} \rr 
}
\sum_{\, \bY}
\bW^{         }_{\, \bfq \, \bY_1 / \bY} \ll y^{\, \bj -1}\, x^{-  Y_3} \rr 
\bW^{\, \prime}_{\, \bfq \, \bY_2 / \bY} \ll x^{\, \bi   }\, y^{ - Y_3^{\, \prime} } \rr, 
\label{macdonald.vertex}
\end{empheq}

$\bi = \ll 1, 2, \cdots \rr$, and the arguments in 
$\bW^{         }_{\, \bY_1 / Y} \ll y^{\, \bj -1}\, x^{-Y_3} \rr$ and
$\bW^{\, \prime}_{\, q \,  \bY_2 / Y} \ll x^{\, \bi   }\, y^{-Y_3^{\, \prime}   } \rr$
are in the sense of Section \ref{sequences}. 

\subsection{The twisted version of the vertex}

We define the twisted version of the vertex $\cE^{\, \star}$, in the same way that 
we defined $\cE$, but with the choice of arguments in Step \textbf{2} changed to,  

\begin{equation}
\phi_{\, q \, +} \ll x^{\, - \, j \, + \, 1}\, y^{\,      y^{\, \prime}_{\, 3, \, j}} \rr, 
\quad 
\phi_{\, q \, -} \ll y^{\, i               }\, x^{\, - \, y^{         }_{\, 3, \, i}} \rr, 
\label{full.arguments.twisted}
\end{equation}

and calculate the expectation value in the twisted basis, Step \textbf{3}, as
\footnote{\, 
We can alternatively use the vertex operators, 
$\bi \, . \, \phi_{\, q \, +} \ll  x^{-j + 1}\, y^{ y^{\, \prime}_{3, j}}  \rr$, 
and  
$\bi \, . \, \phi_{\, q \, -} \ll y^{   i}\, x^{  -y_{3, i}            }   \rr$, 
and the usual $q$-Whittaker basis to obtain the same result. 
}
,  

\begin{equation}
\cE^{\, \star \, unnorm}_{\, \bY_1 \, \bY_2 \, Y_3} \ll x, y \rr 
=
\langle\, 
\bW^{\, \star}_{\, \bfq \, \bY_1} 
\, |\,
\ll 
\prod_{ \textit{ Maya} \ll Y_3 \rr}
\phi_{\, q \, \pm} 
\rr 
\, |\, 
\bW^{\, \prime\, \star}_{\, \bfq \, \bY_2} \rangle
\label{unnormalized.elliptic.vertex.twisted}
\end{equation}

A parallel calculation leads to the following 
expression for the twisted version of the elliptic vertex,

\begin{empheq}[box=\fbox]{equation}
\cE^{\, \star}_{\, \bY_1 \, \bY_2 \, Y_3} \ll x, y \rr 
= 
\prod_{\wsq \in Y_3}
\frac{
1
}{
\theta_q \ll \,x^{\,L_{\wsq \, Y_3}}\, y^{\,A^{\, +}_{\wsq \, Y_3}} \rr 
}
\sum_{\, \bY}
\bW^{\,          \star}_{\, \bfq \, \bY_1 / \bY} \ll y^{\, \bi     }\, x^{-  Y_3}              \rr 
\bW^{\, \prime\, \star}_{\, \bfq \, \bY_2 / \bY} \ll x^{\, \bj -1  }\, y^{ - Y_3^{\, \prime} } \rr
\label{macdonald.vertex.twisted}
\end{empheq}

\subsection{The $q \rightarrow 0$ limit}
In the $q \rightarrow 0$ limit, the $q$-Whittaker function $W_{q \,  Y_1/Y_2}(x)$ reduces to 
the Schur function $s_{Y_1/Y_2} \ll \bx \rr$. As we can see from the expression of vertex 
operators, equation \ref{gamma.vertex.operators}, 
$\Gamma_{q^{\,\prime} \, b \pm}(x^{\, \prime})$ goes to $1$ in the $q \rightarrow 0$ limit and, 

\begin{equation}
W^{\, \star}_{q^{\, \prime} \, \, Y_1/Y_2}\rightarrow \delta_{Y_1,Y_2},
\end{equation}

in this limit. This trivializes the $q^{\, \prime}$-dependent part of the partition function, 
as long as we consider toric diagrams with trivial external legs. We can effectively drop the 
$q^{\, \prime}$-part in the elliptic vertex when we compute the partition function in the limit. 
Therefore, we can simply replace,  

\begin{equation}
\bW_{\bfq \, \bY / \bR} \ll \bx \rr 
\rightarrow 
s_{Y_1/R_1} \ll \bx \rr,
\quad 
\bW^{\, \prime}_{\bfq \, \bY / \bR} \ll \bx \rr 
\rightarrow 
s_{Y_1/R_1} \ll \bx \rr, 
\quad 
\theta_p \ll x \rr \rightarrow \ll 1-x \rr,
\end{equation}

in this limit, and the elliptic vertex reduces to the refined vertex in \cite{iqbal.kozcaz.vafa}, 

\begin{equation}
\cE^{        }_{\, \bY_1 \, \bY_2 \, Y_3} \ll x, y \rr 
\rightarrow 
\cR_{\, Y^{\, \prime}_{1A}\, Y_{\, 2 \, A}       \, Y_3       } \ll y, x \rr,
\quad 
\cE^{\, \star}_{\, \bY_1 \, \bY_2 \, Y_3} \ll x, y \rr 
\rightarrow 
\cR_{\, Y_{\, 1 \, A}       \, Y^{\, \prime}_{2A}\, Y^{\, \prime}_3} \ll x, y \rr 
\end{equation}

\subsection{Equivalence with the elliptic Awata-Feigin-Shiraishi vertex}\label{s:equiv-AFS}
The vertex operators used to construct $\cE$ are equivalent to those used to construct the 
elliptic Awata-Feigin-Shiraishi vertex constructed in \cite{zhu.01}. To see this, we focus 
on the simple case of $\cE_{\, \bY_1 \, \bY_2 \, \emptyset} \ll x, y \rr$. In this case, 
the corresponding AFS vertex is the normal-ordered product
\footnote{\,
We use the notation in Equation \ref{abbreviation} for infinite products of vertex operators, 
the definition of the two-boson vertex operators in terms of single-boson vertex operators, 
Equation \ref{phi.vertex.operators}, the definition of the single-boson vertex operators, 
Equation \ref{gamma.vertex.operators}.
},

\begin{equation}
\Phi_\emptyset \ll 1 \rr = 
\, : \,
\phi_{\, - \, q} \ll x^{\,   \bj    } \rr \, 
\phi_{\, + \, q} \ll y^{\, - \bi + 1} \rr
\, : \, 
\label{afs.vertex}
\end{equation}

The infinite products of two-boson vertex operators on the right hand side of Equation
\ref{afs.vertex} can be evaluated in the form, 

\begin{multline}
\phi_{\, + \, q} \ll y^{\, - \bi + 1} \rr = 
\exp 
\ll 
\sum_{n=1}^{\infty} \,
\frac{
1
}{
n
}
\frac{
1
}{
\ll 1 - y^{\, n} \rr \, 
\ll 1 - q^{\, n} \rr
} \, 
a_{n}
\rr 
\exp 
\ll 
\sum_{n=1}^{\infty} \,
\frac{
1
}{
n
} 
\frac{
1
}{
\ll 1 - y^{\, \prime \, n} \rr \, 
\ll 1 - q^{\, \prime \, n} \rr
} 
\, 
b_n
\rr,
\\
\phi_{\, - \, q} \ll x^{\, \bj} \rr = 
\exp 
\ll 
\, - \, 
\sum_{n=1}^{\infty} \,
\frac{
1
}{
n
}
\frac{
1
}{
\ll 1 - x^{\, \prime \, n} \rr 
\ll 1 - q^{\, n} \rr
}
\, 
a_{-n}
\rr 
\exp 
\ll 
\, - \, 
\sum_{n=1}^{\infty} \,
\frac{
1
}{
n
}
\frac{
1
}{
\ll 1 - x^{\,        \, n} \rr  
\ll 1 - q^{\, \prime \, n} \rr
} 
\, 
b_{-n}
\rr
\end{multline}

Setting, 

\begin{multline}
a_n = a^{\, Saito}_n,
\quad 
a_{-n} = 
\ll 
\frac{
1 - x^{\, \prime \,  n}
}{
1 - y^{\, \prime \,  n} 
}
\rr 
a^{\, Saito}_{\, - n},
\quad 
n = 1, 2, \cdots,
\\
b_n = - b^{\, Saito}_n,
\quad 
b_{\, - n} = 
- 
\ll 
\frac{
1 - x^{\, n} 
}{
1 - y^{\, n}  
}
\rr \, 
b^{\, Saito}_{\, - n},
\quad 
n = 1, 2, \cdots,
\label{redef}  
\end{multline}

where $a^{\, Saito}_n$ and $b^{\, Saito}_n$ are the $pqt$-Heisenberg generators, 
Equation \ref{two.heisenbergs}, 
used in \cite{saito.01, saito.02, saito.03} and in \cite{zhu.01}, 
with the parameters reset as,

\begin{equation}
p \rightarrow q,             \quad
q \rightarrow y^{\, \prime}, \quad
t \rightarrow x^{\, \prime},
\end{equation}

where the parameters on the left are Saito's and the parameters on the right
are those used in the present work, we obtain 

\begin{multline}
\Phi_\emptyset \ll 1 \rr =
\\ 
: \, 
\exp \ll  \sum_{n \neq 0}
\frac{
1
}{
n
} \, 
\frac{
1
}{
\ll 1 - y^{\, n} \rr 
\ll 1 - q^{\, | \, n \, |} \rr
}
\,
a^{\, Saito}_{n}
\rr 
\, 
\exp 
\ll
\, - \, 
\sum_{n \neq 0} 
\frac{
1
}{
n
} \, 
\frac{
1
}{
\ll 1 - y^{\, \prime \, n} \rr 
\ll 1 - q^{\, \prime \, | \, n \, |} \rr
} \, 
b^{\, Saito}_n
\rr
\, :
\end{multline}

which is the correct expression for the elliptic AFS vertex for $Y_3 = \emptyset$ \cite{zhu.01}. 
With the mapping in Equation \ref{redef}, one can repeat the argument in \cite{awata.feigin.shiraishi} 
to prove the equivalence. We do not do this here as it is not the main point of this paper. 

\section{The 6D strip partition function}
\label{section.10}
\textit{We compute the 4-vertex strip partition function obtained by gluing four elliptic 
vertices, and show that the result is a 6D strip partition function.}

\subsection{The rules of gluing}
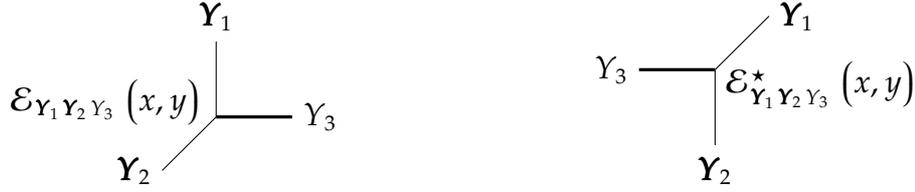
\begin{figure}
\centering
\begin{tikzpicture}
\draw [very thick] (0,0)--(1,0);
\draw (0,0)--(0,1);
\draw (0,0)--(-0.71,-0.71);
\node at (1,0) [right] {$Y_3$};
\node at (0,1) [above] {$\bY_1$};
\node at (-0.71,-0.71) [left] {$\bY_2$};
\node at (0,0.2) [left] {$\cE_{\, \bY_1 \, \bY_2 \, Y_3} \ll x, y \rr$};
\end{tikzpicture}
\hskip 3cm
\begin{tikzpicture}
\draw (0,0)--(0.71,0.71);
\draw [very thick] (0,0)--(-1,0);
\draw (0,0)--(0,-1);
\node at (0,-1) [below] {$\bY_2$};
\node at (-1,0) [left] {$Y_3$};
\node at (0.71,0.71) [right] {$\bY_1$};
\node at (0,-0.2) [right] {$\cE^{\, \star}_{\, \bY_1 \, \bY_2 \, Y_3} \ll x, y \rr $};
\end{tikzpicture}
\caption{The assignment of the elliptic vertex and its twisted version to toric diagrams.
The preferred legs are shown as thick horizontal lines.}
\label{fig:assign-vertex}
\end{figure}

Drawing web diagrams, we set all horizontal legs to be the preferred. Having done that, 
the set of all vertices splits into into two disjoint subsets, 
one       with all preferred legs pointing to the right, and 
the other with all preferred legs pointing to the left, as in Figure \ref{fig:assign-vertex}. 
We take the first subset to consist of elliptic vertices, and 
the second to consist of twisted vertices. 
We glue vertices only to twisted vertices and \textit{vice versa}. 

\subsubsection{The choice of K\"ahler parameters}
\label{choice.kahler.parameters}
We form strips by gluing elliptic vertex non-preferred legs. Each of these 
elliptic vertex non-preferred legs is effectively a pair of refined vertex 
non-preferred legs, and each of the latter is assigned a Young diagram. 
Since the assigned Young diagrams are independent, we can in principle use 
a pair of independent K\"ahler parameters $\ll Q_{\, A}, Q_{\, B} \rr$ to 
glue. 
However, to recover the 6D partition functions computed in the literature, 
including \cite{hollowood.iqbal.vafa, iqbal.kozcaz.yau, nieri}, we need to 
choose the K\"ahler parameter pair as $\ll Q, Q^{\, \prime} \rr$, where 
$Q^{\, \prime} = 1 / Q$, consistently with all choices of all parameter 
pairs made so far. At this stage, we have no justification for this choice 
other than reproducing the known 6D partition functions.

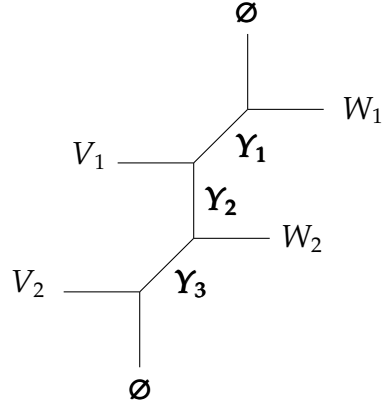
\begin{figure}
\begin{center}
\begin{tikzpicture}
\draw (0,0)--(0.71,0.71);
\draw (0,0)--(-1,0);
\draw (0,0)--(0,-1);
\draw (0.71,0.71)--(0.71,1.71);
\draw (0.71,0.71)--(1.71,0.71);
\draw (0,-1)--(1,-1);
\draw (0,-1)--(-0.71,-1.71);
\draw (-0.71,-1.71)--(-1.71,-1.71);
\draw (-0.71,-1.71)--(-0.71,-2.71);
\node at ( 0.40, 2.00) [right] {$\pmb{\emptyset}$};
\node at ( 1.80, 0.70) [right] {$W_1$};
\node at ( 0.40, 0.20) [right] {$\pmb{Y_1}$};
\node at (-1.00, 0.10) [left]  {$V_1$};
\node at ( 0.00, -0.50) [right] {$\pmb{Y_2}$};
\node at ( 1.00, -1.00) [right] {$W_2$};
\node at (-0.40,-1.60) [right] {$\pmb{Y_3}$};
\node at (-1.80,-1.60) [left]  {$V_2$};
\node at (-1.00,-3.00) [right] {$\pmb{\emptyset}$};
\end{tikzpicture}
\end{center}
\caption{The strip obtained by gluing 4 vertices along their non-preferred legs.
$V_1$, $V_2$, $W_1$ and $W_2$ are single diagrams. 
$\pmb{Y_1}$, $\pmb{Y_2}$ and $\pmb{Y_3}$ are pairs of Young diagrams. 
$\pmb{\emptyset}$ is a pair of empty Young diagrams.}
\label{fig:strip-2}
\end{figure}

\subsection{Example. The 4-vertex strip partition functions}

The partition function of the 4-vertex strip in Figure \ref{fig:strip-2} is, 

\begin{multline}
Z_{\, 4-strip} =
\sum_{\bY_{\, 1, \, 2, \, 3}} 
\prod_{\ell = 1}^3 
\ll - Q_\ell \rr^{| \bY_\ell |}
\cE^{        }_{\, \pmb{\emptyset} \, \bY_1           \, W_1} \ll x, y \rr
\cE^{\, \star}_{\,      \bY_1      \, \bY_2           \, V_1} \ll x, y \rr
\cE^{        }_{\,      \bY_2      \, \bY_3           \, W_2} \ll x, y \rr
\cE^{\, \star}_{\,      \bY_3      \, \pmb{\emptyset} \, V_2} \ll x, y \rr
\\
=
\prod_{k = 1}^2
\frac{
1
}{
\theta_q \ll x^{\, L^{\, +}_{\wsq, \, W_k}} \, y^{\, A^{    }_{\wsq, \, W_k}} \rr 
\theta_q \ll x^{\, L^{    }_{\wsq, \, V_k}} \, y^{\, A^{\, +}_{\wsq, \, V_k}} \rr 
} 
\, 
\sum_{\bY_{1, 2, 3}}
\sum_{\bR_{1,2}}
\prod_{\ell =1}^3 \ll - Q_\ell \rr^{| \bY_\ell |} 
\\ 
\bW^{\, \prime         }_{\, \bfq \, \bY_1      } \ll x^{\, \bi} y^{\, - W^{\, \prime}_1} \rr
\ll 
\bW^{\,  \star         }_{\, \bfq \, \bY_1/\bR_1} \ll y^{\, \bj} x^{\, - V^{         }_1} \rr \, 
\bW^{\, \prime \, \star}_{\, \bfq \, \bY_2 / \bR_1}  \ll x^{\, \bi - 1} y^{\,  - V^{\, \prime}_1}  \rr 
\rr 
\\
\ll 
\bW^{                  }_{\, \bfq \,        \bY_2 / \bR_2}  \ll y^{\, \bi - 1} x^{\,  - W^{         }_2}  \rr \, 
\bW^{\, \prime         }_{\, \bfq \,        \bY_3 / \bR_2}  \ll x^{\, \bj    } y^{\,  - W^{\, \prime}_2}  \rr 
\rr \, 
\bW^{\, \star          }_{\, \bfq \,        \bY_3        }  \ll y^{\, \bj    } x^{\,  - V_2            }  \rr 
\label{4.strip}
\end{multline}

where $|\bY| = |Y_A| - |Y_B|$ because of our choice of K\"ahler parameters, see section
\ref{choice.kahler.parameters}. 

\subsection{Using the Cauchy identities}
We compute the strip partition function in Equation \ref{4.strip} in 4 steps using the Cauchy identities
derived in section \ref{section.07}.

\subsubsection{Step 1} 
Using the identity
$Q^{\, | \bY | \, - \, | \bX |} \, \bW_{\bY / \bX} \, \ll \bx \rr = 
\bW_{\bY / \bX} \, \ll Q \, \bx \rr$, 
where $\bX$ and $\bY$ are partition pairs, and $\bx$ is a set of variables, 
which follows from the properties of the $q$-Whittaker functions 
that make $\bW$, and the Cauchy identity \ref{double-star-normal-Cauchy} 
to perform the sum over $\bY_1$ and over $\bY_3$
\footnote{\,
To relate the notation $\bY_1$ and $\bY_3$ for the Young diagrams that are summed 
out to that of the Young diagrams that are introduced, 
we use $\bR_1$ and $\bR_3$ for the latter, and there is no $\bR_2$
}, 

\begin{multline}
Z_{\, 4-strip} = 
\prod_{k = 1}^2
\frac{
1
}{
\theta_q \ll x^{\, L^{\, +}_{\, \wsq, \, W_k}} \, y^{\, A^{    }_{\, \wsq, \, W_k}} \rr 
\theta_q \ll x^{\, L^{    }_{\, \wsq, \, V_k}} \, y^{\, A^{\, +}_{\, \wsq, \, V_k}} \rr 
}
\, 
\sum_{\bY_2} 
\sum_{\bR_1 \, \bR_3}
\ll - Q_2 \rr^{| \bY_2 |} \, \times
\\ 
\prod_{i, j = 1}^\infty 
\theta_q \ll Q_1 \, x^{\, i - V_{\, 1, \, j}} \, y^{\, j- W^{\, \prime}_{\, 1, \, i}} \rr  
\bW^{\, \prime}_{\, \bfq \, \bR_1} \ll - Q_1 \, x^{\, \bi} \, y^{\, - W^{\, \prime}_1} \rr \,
\bW^{\, \prime \, \star}_{\, \bfq \, \bY_2 / \bR_1}  \ll x^{\, \bi - 1} y^{\,  - V^{\, \prime}_1}  \rr 
\\ 
\bW^{                  }_{\, \bfq \, \bY_2 / \bR_3}  \ll y^{\, \bi - 1} x^{\,  - W^{         }_2}  \rr 
\, 
\prod_{i, j = 1}^\infty 
\theta_q \ll Q_3 \, x^{\, i - V_{2, \, j}} \, y^{\, j - W^{\, \prime}_{2, \, i}} \rr   
\bW^{\, \star }_{\, \bfq \, \bR_3}  \ll - Q_3 \, y^{\, \bj} \, x^{\, - \, V_2} \rr
\\
= 
\prod_{k = 1}^2
\frac{
1
}{
\theta_q \ll x^{\, L^{\, +}_{\, \wsq, \, W_k}} \, y^{\, A^{    }_{\, \wsq, \, W_k}} \rr 
\theta_q \ll x^{\, L^{    }_{\, \wsq, \, V_k}} \, y^{\, A^{\, +}_{\, \wsq, \, V_k}} \rr 
}
\\
\prod_{i, j = 1}^\infty 
\theta_q \ll Q_1 \, x^{\, i - V_{\, 1, \, j}} \, y^{\, j- W^{\, \prime}_{\, 1, \, i}} \rr \, 
\prod_{i, j = 1}^\infty 
\theta_q \ll Q_3 \, x^{\, i - V_{2, \, j}} \, y^{\, j - W^{\, \prime}_{2, \, i}} \rr  
\\  
\sum_{\bY_2} 
\sum_{\bR_1 \, \bR_3}
\ll - Q_2 \rr^{| \bY_2 |} \,
\bW^{\, \prime}_{\, \bfq \, \bR_1} \ll - Q_1 \, x^{\, \bi} \, y^{\, - W^{\, \prime}_1} \rr \,
\bW^{\, \prime \, \star}_{\, \bfq \, \bY_2 / \bR_1}  \ll x^{\, \bi - 1} y^{\,  - V^{\, \prime}_1}  \rr 
\\ 
\bW^{                  }_{\, \bfq \, \bY_2 / \bR_3}  \ll y^{\, \bi - 1} x^{\,  - W^{         }_2}  \rr 
\, 
\bW^{\, \star }_{\, \bfq \, \bR_3}  \ll - Q_3 \, y^{\, \bj} \, x^{\, - \, V_2} \rr
\end{multline}

\subsubsection{Step 2}
Using the Cauchy identity \ref{double-normal-star-Cauchy} to perform the sum over $\bY_2$, 

\begin{multline}
Z_{\, 4-strip} = 
\prod_{k = 1}^2
\frac{
1
}{
\theta_q \ll x^{\, L^{\, +}_{\wsq, \, W_k}} \, y^{\, A^{    }_{\, \wsq, \, W_k}} \rr 
\theta_q \ll x^{\, L^{    }_{\wsq, \, V_k}} \, y^{\, A^{\, +}_{\, \wsq, \, V_k}} \rr 
}
\\ 
\prod_{i, j = 1}^\infty 
\theta_q \ll Q_1 \, x^{\, i - V_{\, 1, \, j}} \, y^{\, j - W^{\, \prime}_{1, \, i}} \rr \,
\prod_{i, j = 1}^\infty 
\theta_q \ll Q_2 x^{\, i - 1 - W_{\, 2, \, j}} y^{\, j - 1 - V^{\, \prime}_{\, 1, \, i}} \rr \,
\prod_{i, j = 1}^\infty 
\theta_q \ll Q_3 \, x^{\, i - V_{2, \, j}} \, y^{\, j - W^{\, \prime}_{2, \, i}} \rr
\\
\sum_{\bS} \,  
\sum_{\bR_{\, 1, \, 3}} \, 
\ll - Q_2 \rr^{| \bR_1 | } \, 
\bW^{\, \prime}_{\, \bfq \, \bR_1} \ll - Q_1 \, x^{\, \bi} \, y^{\, - W^{\, \prime}_1} \rr \, 
\bW^{                  }_{\, \bfq \, \bR_1 / \bS} \ll          y^{\, \bi - 1} \, x^{\, - W^{         }_2} \rr
\\ 
\bW^{\, \prime \, \star}_{\, \bfq \, \bR_3 / \bS} \ll - Q_2 \, x^{\, \bi - 1} \, y^{\, - V^{\, \prime}_1} \rr \, 
\bW^{\, \star }_{\, \bfq \, \bR_3}  \ll - Q_3 \, y^{\, \bj} \, x^{\, - \, V_2} \rr, 
\end{multline}

\subsubsection{Step 3}
Using the Cauchy identities \ref{double-normal-normal-Cauchy} and \ref{double-star-star-Cauchy} 
to sum over $\bR_1$ and over $\bR_3$, 

\begin{multline}
Z_{\, 4-strip} = 
\prod_{k = 1}^2
\frac{
1
}{
\theta_q \ll x^{\, L^{\, +}_{\wsq, \, W_k}} \, y^{\, A^{    }_{\wsq, \, W_k}} \rr 
\theta_q \ll x^{\, L^{    }_{\wsq, \, V_k}} \, y^{\, A^{\, +}_{\wsq, \, V_k}} \rr 
}
\\
\frac{   
\prod_{i, j = 1}^\infty 
\theta_q \ll Q_1 \, x^{\, i - V_{1, \, j}} \, y^{\, j - W^{\, \prime}_{1, \, i}} \rr \,
\prod_{i, j = 1}^\infty 
\theta_q \ll Q_{\, 2     } \, x^{\, i - 1 - W_{\, 2, \, j}} \, y^{\, j - 1 - V^{\, \prime}_{\, 1, \, i}} \rr \, 
\prod_{i, j = 1}^\infty 
\theta_q \ll Q_3 \, x^{\, i - V_{\, 2, \, j}} \, y^{\, j - W^{\, \prime}_{\, 2, \, i}} \rr
}{
\prod_{i, j = 1}^\infty
\theta_q \ll Q_{\, 1 \, 2} \, x^{i - W_{\, 2, \, j}} \, y^{j - 1 - W^\prime_{\, 1, \, i}} \rr \, 
\prod_{i, j = 1}^\infty
\theta_q \ll Q_{\, 2 \, 3} \, x^{\, i - 1 - V_{\, 2, \, j}} \, y^{\, j -     V^{\, \prime}_{\, 1, \, i}} \rr
}
\\ 
\sum_{\bS} \, 
\bW^{\, \prime}_{\, \bfq \, \bS} \ll  Q_{\, 1 \, 2} \, x^{\, \bi} \, y^{\, - W^{\, \prime}_1} \rr 
\,  
\bW^{\, \star }_{\, \bfq \, \bS}  \ll - Q_{23} \, y^{\, \bj} \, x^{\, - \, V_2} \rr, 
\end{multline}

where $Q_{i j} = Q_i \, Q_j$

\subsubsection{Step 4}
Using the Cauchy identity \ref{double-star-normal-Cauchy} to perform the sum over $\bS$, 

\begin{multline}
Z_{\, 4-strip} =
\prod_{k = 1}^2 
\frac{
1
}{
\theta_q \ll x^{\, L^{\, +}_{\wsq, \, W_k}} \, y^{\, A^{\,  }_{\wsq, \, W_k}} \rr
\theta_q \ll x^{\, L^{    }_{\wsq, \, V_k}} \, y^{\, A^{\, +}_{\wsq, \, V_k}} \rr
}
\\ 
\frac{
N^{\, \prime}_{\, W_1 \, V_1} \ll Q_1    \,  y                                                \rr 
N^{\, \prime}_{\, V_1 \, W_2} \ll Q_2    \,  x^{\, \prime}                                    \rr 
N^{\, \prime}_{\, W_2 \, V_2} \ll Q_3    \,  y                                                \rr
}{
N^{\, \prime}_{\, W_1 \, W_2} \ll Q_{\, 1 \, 2}     \rr 
N^{\, \prime}_{\, V_1 \, V_2} \ll Q_{\, 2 \, 3} y \, x^{\, \prime} \rr 
} \, 
N^{\, \prime}_{\, W_1 \, V_2} \ll Q_{\, 1 \, 2 \, 3}\, y \rr, 
\\ 
N^{\, \prime}_{\, Y_1 \, Y_2} \ll Q \rr 
= \prod_{i, j = 1}^\infty
\theta_q 
\ll 
Q \, 
y^{ - Y^{\, \prime}_{1,j} + i - 1} \, 
x^{-Y_{2,i}       + j}
\rr, \, 
Q_{\, 1 \, 2 \, 3} = 
Q_1 \, Q_2 \, Q_3,  
\label{partition-function-comp}
\end{multline} 

which agrees with that in \cite{hollowood.iqbal.vafa, iqbal.kozcaz.yau, nieri}.

\subsubsection{Remark} The product 
$N^{\, \prime}_{\, Y_1 \, Y_2} \ll Q \rr$ in Equation \ref{partition-function-comp} 
is $N^{\,       }_{\, Y_1 \, Y_2} \ll Q \, | \, q_1, q_2, q \rr$,
$q_1 = y^{\, \prime}, q_2 = x$,
in \cite{awata.kanno.02}, up to a factor, 

\begin{equation}
N^{\, \prime}_{\, Y_1 \, Y_2} \ll Q \rr =
\ll 
\prod_{i,j=1}^\infty 
\theta_q \ll Q \, y^{\, i - 1} \, x^{\, j} \rr 
\rr \, 
N^{         }_{\, Y_1 \, Y_2} \ll Q \, | \, y^{\, \prime}, x, q \rr 
\end{equation}

\subsubsection{Remark} 
In computing the 6D instanton partition function in Equation \ref{partition-function-comp}, 
we used the variables $x^{\, \bi} y^{\, - \, Y^{\, \prime}_3}$ and 
$y^{\, \bj - 1} \, x^{\, -Y_3}$ in Equation \ref{macdonald.vertex} instead of the variables  
$x^{\, -\rho} \, y^{-Y^{\, \prime}_3}$ and $y^{\, - \, \rho} \, x^{\, -Y_3}$ in
\cite{iqbal.kozcaz.vafa}, where
$x^{\, - \, \rho} \, y^{\, Y} = 
\ll 
x^{\, 1/2} \, y^{\, y_1}, \, 
x^{\, 3/2} \, y^{\, y_2}, \, 
x^{\, 5/2} \, y^{\, y_3}, \cdots \rr$. 
To compensate for these differences when comparing our results with those obtained using 
the refined topological vertex of \cite{iqbal.kozcaz.vafa}, we need to rewrite the K\"ahler
parameters in Equation \ref{partition-function-comp} as, 

\begin{equation}
Q_{\, 2 i    } = \ll y  x^{\, \prime} \rr^{ 1/2} \, Q_{\, 2 i    }^{\, IKV}, 
\quad
Q_{\, 2 i - 1} = \ll x  y^{\, \prime} \rr^{ 1/2} \, Q_{\, 2 i - 1}^{\, IKV},
\quad
i = 1, 2, \cdots, 
\end{equation}

where $Q_{\, i}^{\, IKV}, i = 1, 2, \cdots,$ is identified with the corresponding K\"ahler parameter 
in \cite{iqbal.kozcaz.vafa}.  

\subsubsection{More general strip partition functions}
Strip partition functions with more vertices can be calculated, using the same Cauchy identities as in 
the 4-vertex case. The result of these computations is that the parition function of an $N$-vertex strip,
$N=6, 8, \cdots,$ computed using elliptic topological vertices, with empty top and bottom external legs, 
is equal to the parition function of the same strip computed using refined topological vertices,with the 
top and bottom legs identified. Gluing copies of these strips, computed either way, along their horizontal 
preferred legs, we obtain 6D instanton partition function. 

\section{Comments}
\label{section.11}

\subsection{The $q \rightarrow 0$ limit}
In the limit $q \rightarrow 0$, all $\theta_q \ll x \rr \rightarrow 1-x$, and the 6D partition function in Equation \ref{partition-function-comp} 
reduces to the 5D partition function of the corresponding toric diagram computed using the refined topological 
vertex. From the viewpoint of compactification, $q\rightarrow 0$ corresponds to the length of the 
compactification circle going to infinity, which forces the vertical external legs to be trivial. 

\subsection{The $R \rightarrow 0$ limit}
Another interesting limit is obtained by taking the radius of the 
$M$-theory circle $R \rightarrow 0$, while keeping $q$ finite, to 
obtain a 5D partition function \cite{foda.gavrylenko}. A study of 
this 5D partition function, its relation to that obtained in the 
limit $q \rightarrow 0$, and the possible interplay of these two 
limits is beyond the scope of the present work.

\subsection{The other involution and other Cauchy identities} 
In addition to the Cauchy identities that involve $q$-Whittaker functions and their twisted
versions, there are identities that involve $q$-Whittaker functions and Hall-Littlewood 
functions. 
These identities are obtained by the action of the Macdonald involution $\omega$, Equation
\ref{involution.m}, that changes the $q$-Whittaker functions labelled by a Young diagram $Y$ 
and a parameter $q$ to a Hall-Littlewood function labelled by the transpose diagram 
$Y^{\, \prime}$ and the same parameter $q$ \cite{borodin.wheeler}.
We did not consider $\omega$ and the resulting Cauchy identities because they do not lead 
to the 6D instanton partition functions that we wish to compute. Instead, we considered 
the involution $\bi$, Equation \ref{involution} that leads to the twisted $q$-Whittaker 
functions and the Cauchy identities that lead to the 6D instanton partition functions. 
In this sense, our construction of $\cE$ aimed in a specific direction and is not the only
construction possible. It is not clear what the construction that uses $\omega$ gives
\footnote{\,
One can make a similar remark regarding our choice of the K\"ahler parameters in 
section \ref{choice.kahler.parameters}, which was motivated by producing the 6D partition 
functions in the literature, including \cite{hollowood.iqbal.vafa, iqbal.kozcaz.yau, nieri}, 
and only that.
}.

\subsection{The 2D interpretation}
In addition to their interpretation as 6D instanton partition functions, the partition 
functions obtained by gluing copies of $\cE$ and $\cE^{\, \star}$ have a natural 
interpretation as elliptic deformations of 2D conformal blocks. 
We expect that these 2D elliptic conformal blocks are related to the $n$-point local height 
probabilities in off-critical exactly-solved statistical mechanical models, with elliptic 
Boltzmann weights, studied in \cite{jimbo.miki.miwa.nakayashiki, lukyanov.pugai}. 

\subsection{Two extensions that need physical interpretation}
\subsubsection{Both methods can work in parallel}
Using the elliptic vertex can be combined with the trace method. 
One can compute an elliptic partition function using the elliptic vertex, then take the trace over 
the all possible external states on the doubled non-preferred legs. It is not clear what the result 
means.

\subsubsection{The Macdonald parameter $t$} One advantage of using $\cE$, as opposed to taking traces 
as in \cite{hollowood.iqbal.vafa} is that it makes it obvious that there is room for one parameter, 
namely the second Macdonald parameter $t$. We have not switched this parameter on because it does not 
appear in the 6D instanton partition function results of \cite{hollowood.iqbal.vafa}. We could have 
easily switched $t$ on, but we would have no interpretation for what that means
\footnote{\,
In \cite{sulkowski}, Sulkowski showed that starting from the topological string partition function on 
$\cC^3$ then switching on the Macdonald parameter $t$ produces the topological string partition function 
of the conifold, with the parameter $t$ parameterizing the size of the sphere $P^1$. We expect that introducing $t$ in more general topological string partition functions 
will have a related effect.
}. 

\subsection{The Clavelli-Shapiro trace reduction method}
In \cite{clavelli.shapiro}, Clavelli and Shapiro proposed a method to reduce the evaluation of 
a trace of exponentials of Heisenberg generators, of the type that appears in string theory and 
in the present work, to the evaluation of a single vacuum expectation value of exponentials of 
\textit{two} Heisenberg generators
\footnote{\,
Apart from a slight change of notation, this outline follows section \textbf{C.1}, p. 522, of 
\cite{clavelli.shapiro}.
}. 
Given a single Heisenberg annihilation operator $a$ and its conjugate creation operator $a^{\, \dagger}$, 
a product $\cO \ll a, a^{\, \dagger} \rr$ of exponentials in $a$ and $a^{\, \dagger}$, and a parameter 
$x < 1$, we wish to evaluate the trace,

\begin{equation}
Tr \ll \, x^{\, a^{\, \dagger} \, a} \cO \ll a, a^{\, \dagger} \rr \rr = 
\sum_{n = 0}^\infty \,  
\langle \, n \, | \, 
x^{\, a^{\, \dagger} \, a} \, \cO \, \ll a, a^{\, \dagger} \rr \, 
| \, n \, \rangle, 
\quad
[a, a^{\, \dagger}] = 1,
\label{trace.heisenberg.01}
\end{equation}

where $\langle \, n \, |$ and $| \, n \, \rangle$ are the state created by the action of $n$ copies of $a$ 
and of $a^{\, \dagger}$ on the vacuum states $\langle \, 0 \, |$ and $| \, 0 \, \rangle$ respectively.
In \cite{clavelli.shapiro}, Clavelli and Shapiro noticed that by introducing a second Heisenberg algebra, 
generated by $b$ and $b^{\, \dagger}$, that commutes with the first, generated by $a$ and $a^{\, \dagger}$, 
the infinite sum over states on the right hand side of the first Equation \ref{trace.heisenberg.01} becomes,

\begin{equation}
\sum_{n = 0}^\infty \,  
\langle \, n \, | \, 
x^{\, a^{\, \dagger} \, a} \, \cO \, \ll a, a^{\, \dagger} \rr \, 
| \, n \, \rangle 
= 
\langle \, 0 \, | \,
e^{\, b \, a}
x^{\, a^{\, \dagger} \, a} \, \cO \, \ll a, a^{\, \dagger} \rr \,
e^{\, a^{\, \dagger} \, b^{\, \dagger}}
| \, 0 \, \rangle, 
\label{trace.heisenberg.02}
\end{equation}

and the trace has been reduced to computing a single expectation value of operators in a pair of Heisenberg
algebras. Following that, Clavelli and Shapiro show that the right hand side of Equation \ref{trace.heisenberg.02}
can be written in the form,

\begin{multline}
\langle \, 0 \, | \,
e^{\, b \, a}
x^{\, a^{\, \dagger} \, a} \, \cO \, \ll a, a^{\, \dagger} \rr \,
e^{\, a^{\, \dagger} \, b^{\, \dagger}}
| \, 0 \, \rangle 
=
\frac{
1
}{
1 - x
} \, 
\langle \, 0 \, | \,
\cO \ll c, c^{\, \dagger} \rr \, 
| \, 0 \, \rangle, 
\\ 
c = 
\frac{
a
}{
1 - x
}
+
b^{\, \dagger}, 
\quad  
c^{\, \dagger} 
= 
a^{\, \dagger}
-
\frac{
b
}{
1 - 1 / x
}, 
\label{trace.heisenberg.03}
\end{multline}

and the trace of an operator that depends on a single Heisenberg algebra, and a weight (propagator-type) 
parameter $x$ is reduced to a vacuum expectation value of the same operator that now depends on two 
Heisenberg algebras that are deformed in a specific way using the parameter $x$. The above single-mode
result readily extends to the case of a Heisenberg algebra with infinitely-many modes
\footnote{\,
Equation C.4, page 522, in \cite{clavelli.shapiro}.
}, and the conclusion is that traces over exponentials of free fields can be re-written as vacuum 
expectation values in twice the number of free fields. 

\subsubsection{The elliptic vertex \textit{versus} taking traces, and similarities with 
the Clavelli-Shapiro method}
\label{similarities}
The relation between the result in terms of a trace and the same result in terms of no trace but 
twice the number of fields is identical to the relation between the computation of the 6D instanton 
partition functions in terms of the refined vertex and taking traces in \cite{hollowood.iqbal.vafa}, 
and the computation of the same objects in terms of the elliptic vertex proposed in the present work, 
which is basically a deformation of a doubling of the refined vertex. Even the deformation of the pair 
of Heisenberg algebras in Equation \ref{trace.heisenberg.03} is identical to that that appears in the 
present work, and in so far as these computations are concerned, using the elliptic vertex as in the 
present work is related to taking traces as in \cite{iqbal.kozcaz.yau, nieri} \textit{via} 
a Clavelli-Shapiro trace reduction. By using the elliptic vertex, the effect of the compactification 
is local and can be traced to the deformation of each vertex in the strip, unlike in the case of taking 
traces. For that reason, it is possible that, while both methods lead to the same 6D partition functions,
using the elliptic vertex may be more suited to studies of the algebraic structures that underlie 
the elliptic deformation, particularly in 2D integrable models. 

\section*{Acknowledgements}
We thank
F Benini, 
J-E Bourgine,
P Gavrylenko, 
A Hanay, 
C Kozcaz, 
Kimyoung Lee, 
M Manabe, 
Y Matsuo,
V Mitev, 
E Pomoni, 
J Shiraishi, 
Y Tachikawa, 
O Warnaar, 
Jian-Feng Wu, 
F Yagi,
and
G Zafrir
for discussions on this work and on related topics. 
We thank 
the organizers of  
\textit{\lq Combinatorics, Statistical Mechanics, and Conformal Field Theory\rq} 
for hospitality at the Australian Mathematical Research Institute, MATRIX, 
Creswick, Victoria, 
the organizers of 
\textit{\lq Supersymmetric Quantum Field Theories in the Non-perturbative Regime\rq}, 
the Galileo Galilei Institute for Theoretical Physics, Arcetri, Firenze,
and OF thanks
Prof A Dabholkar of the Abdus Salam International Center for Theoretical Physics, 
Trieste, Italy,
Profs K Lechner, M Matone and D Sorokin of the Physics Department, 
University of Padova, Italy
for hospitality at various stages of this work.
OF is supported by a Special Studies Program grant from the Faculty of Science, 
University of Melbourne, and the Australian Research Council. 
RDZ is supported by JSPS fellowship for young students.

\end{document}